\documentclass[11pt]{article}

\usepackage{setspace}
\usepackage{amsmath, amssymb,amsthm,color,comment}
\usepackage{mathrsfs}
\usepackage{bm}
\usepackage{booktabs}
\usepackage[pdftex]{graphicx}

\usepackage{algorithm, algorithmicx, algpseudocode}

\usepackage{float}

\theoremstyle{plain}

\newtheorem{prop}{Proposition}[section]

\theoremstyle{definition}
\newtheorem{rem}{Remark}[section]

\usepackage[authoryear]{natbib}

\makeatletter
 
  \@addtoreset{equation}{section}
\makeatother

\usepackage[top=30mm,bottom=27mm,left=30mm,right=30mm]{geometry}

\title{Approximate Gibbs sampler for Bayesian Huberized lasso
}
\author{Jun Kawakami and Shintaro Hashimoto\\
Department of Mathematics, Hiroshima University, Higashi-Hiroshima, Japan}

  \allowdisplaybreaks
 
\begin{document}

\date{}

\maketitle

\begin{abstract}
The Bayesian lasso is well-known as a Bayesian alternative for Lasso. Although the advantage of the Bayesian lasso is capable of full  probabilistic uncertain quantification for parameters, the corresponding posterior distribution can be sensitive to outliers. To overcome such problem, robust Bayesian regression models have been proposed in recent years. 
In this paper, we consider the robust and efficient estimation for the Bayesian Huberized lasso regression in fully Bayesian perspective. 
A new posterior computation algorithm for the Bayesian Huberized lasso regression is proposed. The proposed approximate Gibbs sampler is based on the approximation of full conditional distribution and it is possible to estimate a tuning parameter for robustness of the pseudo-Huber loss function. Some theoretical properties of the posterior distribution are also derived. We illustrate performance of the proposed method through simulation studies and real data examples.
\end{abstract}

\medskip

\section{Introduction}

Linear regression model is a fundamental tool in modern data analysis. In high-dimensional case, we often use a penalized regression model such as the Lasso (\cite{T96}). The Lasso is to estimate the regression coefficient vector $\bm{\beta}=(\beta_1,\dots,\beta_p)^{\top}$ in the model $\bm{y}=\mu 1_n +X\bm{\beta} +\bm{\varepsilon}$, where $\bm{y}$ is the $n\times 1$ vector of responses, $\mu$ is the intercept, $X$ is the $n\times p$ design matrix, and $\bm{\varepsilon}$ is the $n\times 1$ vector of iid error. The Lasso estimate is achieved by solving the problem
\begin{align}\label{lasso}
\min_{\bm{\beta} \in \mathbb{R}^p} \sum_{i=1}^n (\tilde{y}_i-\bm{x}_i^{\top} \bm{\beta})^2+\lambda \sum_{j=1}^p |\beta_j|,
\end{align}
where $\lambda>0$ is a tuning constant and $\tilde{\bm{y}}=\bm{y}-\bar{y}1_n=(\tilde{y}_1,\dots,\tilde{y}_n)^{\top}$. Efficient optimization methods for Lasso have been developed (e.g. \cite{EHJT04}) and we can easily obtain the point estimate for the Lasso. However, it is well-known that such estimates are often unstable in the presence of outliers. To overcome this problem, robust and sparse linear regression models have been developed in last decade (e.g. \cite{KAZ07}, \cite{WLJ07}, \cite{LZ11}, \cite{ACG13}, \cite{KF17}). Some of these methods are easily implemented by using {\tt robustHD} or {\tt MTE} package in R. However, such frequentist approaches are not capable of full probabilistic uncertainty quantification.

As a Bayesian analog of Lasso, the Bayesian lasso by \cite{PC08} is also well-known. The advantage of the Bayesian approach is capable of uncertainly quantification by using the credible region for example. Furthermore, we can easily estimate a penalization (tuning) parameter from data by using the Gibbs sampling. However, since the original Bayesian lasso is based on the Gaussian likelihood, it suffers from outliers in data. To overcome such problem, some robust Bayesian regression models have been developed in recent years (see e.g. \cite{NR16}, \cite{HS20}, \cite{HIS20}). For example, \cite{HS20} proposed a divergence-based robust posterior distribution to estimate sparse Bayesian linear regression. However, the computation of robust posterior is quite challenging because the posterior distribution is very intractable, and the full conditional distribution for the regression coefficient vector is often non-standard. Although \cite{HS20} proposed a new algorithm which combines sampling and optimization, the computational cost is still not low. Furthermore, the selection of tuning parameter for robustness has been not clear (but recently, some strategies have been developed by \cite{SY21} and \cite{YS21}, for example). 

In classical robust statistics, the Huber loss function (\cite{H64}) is often used, and it is shown that the corresponding estimators have bounded influence function (e.g. \cite{HRRS11}). It is also known that the Huber estimate also has reasonable efficiency for suitable choice of the tuning parameter. In Bayesian approach, \cite{PC08} mentions that robust Bayesian lasso can be implemented by using modified Huber loss function called pseudo-Huber or hyperbolic loss. Beside the Huber-type loss function, the Tukey's biweight loss function is also famous in classical robust statistics. Robust lasso regression based on Tukey's biweight loss was proposed by \cite{CRW18}. Recently, \cite{JR21} also considered the selection of a tuning parameter of the Tukey's biweight loss function using an improper model in Bayesian perspective.

In this paper, we revisit the Bayesian Huberized lasso regression by \cite{PC08} in fully Bayesian perspective. We show some properties of the model (posterior propriety and posterior unimodality) and propose a new posterior computation algorithm based on an approximate Gibbs sampling which includes the selection of the tuning parameter of robustness. The idea comes from \cite{M19}, and we consider an approximation of intractable full conditional distribution by using a gamma distribution. The proposed Gibbs sampler does not depend on some tuning parameters and there exists no rejection steps like the Metropolis-Hastings algorithm. We illustrate the approximation accuracy of full conditional distribution and discuss the robustness of the Bayesian Huberized lasso using influence function through  simulation studies. 

The remaining of the paper is structured as follows: in Section \ref{sec:2}, we formulate the Bayesian Huberized lasso and show some properties. Then, we propose a new Gibbs sampling algorithm which can estimate the tuning parameter of robustness in Section \ref{sec:3}. Robustness properties of the proposed method are also discussed. In Section \ref{sec:4}, the performance of the proposed method compared with existing methods is investigated through simulation. In Section \ref{sec:5}, the proposed Bayesian Huberized lasso model is demonstrated via the analysis of real datasets. In Section \ref{sec:6}, we make a summary and discuss some of future directions. All proofs of propositions are given in Appendix. 

We close this section by introducing some notation used in this article. Let $\mathrm{GIG}(\nu,a,b)$ be the generalized inverse Gaussian (GIG) distribution whose density function is defined by 
\begin{align}\label{GIG1}
f(x)=\frac{(a/b)^{\nu/2}}{2K_\nu(\sqrt{ab})} x^{\nu-1}e^{-\{ax+(b/x)\}/2}\quad(x>0),
\end{align}
where $K_\nu$ is a modified Bessel function of the second kind, $a>0$, $b>0$ and $\nu \in \mathbb{R}$. By setting $\eta=\sqrt{ab}$ and $\rho^2=\sqrt{b/a}$, we can alternatively express the GIG distribution (denoted by $\mathrm{GIG}(\nu,\eta,\rho^2)$) as 
\begin{align}\label{GIG2}
f(x)=\frac{1}{2\rho^2 K_{\nu}(\eta)}\left(\frac{x}{\rho^2}\right)^{\nu-1}e^{-\eta\{(x/\rho^2)+(\rho^2/x) \}/2} \quad (x>0),
\end{align}
where $\eta>0$ is a shape parameter and $\rho^2>0$ is a scale parameter. Let $\mathrm{InvGauss}(\mu,\lambda)$ be the inverse Gaussian distribution whose density function is defined by 
\[f(x)=\sqrt{\frac{\lambda}{2\pi x^3}} \exp\left(-\frac{\lambda(x-\mu)^2}{2\mu^2 x}\right)\quad (x>0),\] 
where $\mu>0$ and $\lambda>0$ are mean and shape parameters, respectively.

\section{Bayesian Huberized lasso}
\label{sec:2}

\subsection{Preliminaries} 
\begin{figure}[htpb]
\centering
\includegraphics[width=15cm]{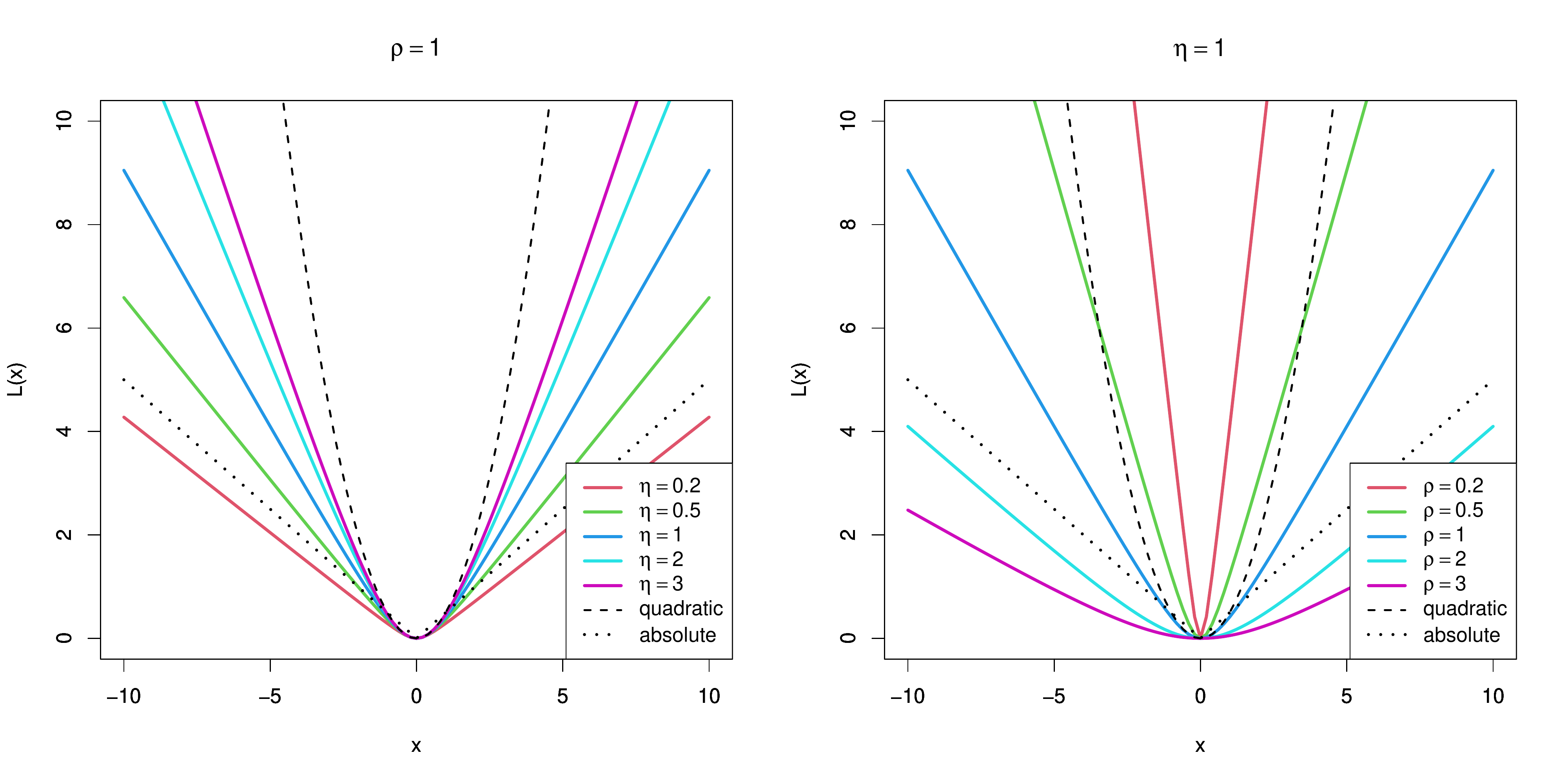}
\caption{Graphs of hyperbolic loss function $L_{\eta,\rho^2}(x)$ compared with quadratic and absolute losses.}
\label{loss_plot}
\end{figure}

In the original lasso problem \eqref{lasso}, if there are outliers in response variables, the parameter estimation based on the squared error $\sum_{i=1}^n (y_i-\bm{x}_i^{\top} \bm{\beta})^2$ in \eqref{lasso} is not robust against outliers. One of remedies for such problem, we can use $\sum_{i=1}^n |y_i-\bm{x}_i^{\top} \bm{\beta}|$ instead of the squared loss function, and the corresponding method is called the least absolute deviation (LAD) regression. However, the LAD regression might be underestimate for non-outlying observations. On the other hand, \cite{RZ04} defined the Huberized lasso regression as
\begin{align*}
\min_{\bm{\beta} \in \mathbb{R}^p} \sum_{i=1}^n L_{c}(y_i-\bm{x}_i^{\top} \bm{\beta})+\lambda \sum_{j=1}^p |\beta_j|,
\end{align*}
where $L_c(\cdot)$ is the Huber loss function defined as 
\begin{align*}
L_c(\bm{y},\bm{x}^{\top} \bm\beta)=
\begin{cases}
(y_i-\bm{x}_i^{\top} \bm{\beta})^{2}   &  \mathrm{if} \ \ |y_i-\bm{x}_i^{\top} \bm{\beta}|\leq c,\\
c\left(|y_i-\bm{x}_i^{\top} \bm{\beta}|-\frac{c}{2}\right)  & \mathrm{otherwise},
\end{cases}
\end{align*}
where $c>0$ is a tuning parameter for robustness which is empirically used as $c=1.345$ (\cite{H64}). 
Since the Huber loss function has nondifferentiable points, some authors use the pseudo-Huber loss function which approximates the Huber loss using differentiable function:
\[L_c^{\mathrm{pH}}(x)=c^2\left( \sqrt{1+\left(\frac{x}{c}\right)^2}-1 \right)=c\sqrt{c^2+x^2}-c^2\]
for $c>0$ (e.g. \cite{HZ03}, \cite{S21}). On the other hand, \cite{PC08} employs  the following hyperbolic loss function to formulate the Bayesian lasso 
\begin{align*}\label{hyperb}
L_{\eta,\rho^2}(x)=\sqrt{\eta (\eta +x^2/\rho^2)}-\eta, 
\end{align*}
where $\eta>0$ is a tuning parameter of robustness and $\rho>0$ is a scale parameter. These two loss functions have the following relation:
\[L_{\eta=c^2,\rho^2=1}(x)=L_c^{\mathrm{pH}}(x).\]
The corresponding density functions $e^{-L(x)}$ for pseudo-Huber and hyperbolic losses are both hyperbolic distribution. Recall the probability density function of the hyperbolic distribution $\mathrm{Hyp}(\mu , \delta , \alpha , \beta)$ given by
\[f(x\mid\mu , \delta , \alpha , \beta)=\dfrac{\gamma}{2\alpha\delta K_{1}(\delta{\gamma})}\exp\{-\alpha\sqrt{\delta^{2}+(x-\mu)^{2}}+\beta(x-\mu)\} \quad (x \in \mathbb{R}),\]
where $\mu,\delta,\alpha,\beta \in \mathbb{R}$. Letting $\beta=0$, $\alpha=\sqrt{\eta/\rho^2}$, $\mu=0$ and $\delta=\sqrt{\eta\rho^2}$, we have $\gamma=\sqrt{\alpha^2-\beta^2}=\sqrt{\eta/\rho^2}$. Hence, the density of the hyperbolic distribution is given by $f(x\mid\mu=0 , \delta=\sqrt{\eta\rho^2} , \alpha=\sqrt{\eta/\rho^2} , \beta=0) \propto \exp\{-L_{\eta,\rho^2}(x)\}$. In Subsection \ref{subsec:2.2}, we formulate the Bayesian Huberized lasso by using the hyperbolic distribution as a working likelihood. We show curves of loss function $L_{\eta,\rho^2}(x)$ in Figure \ref{loss_plot}. We note that for fixed $\rho^2$, $L_{\eta,\rho^2}(x)$ is close to the quadratic loss function as $\eta\to \infty$, and $L_{\eta,\rho^2}(x)$ is close to the absolute loss function as $\eta\to 0$.

\begin{figure}[htpb]
\centering
\includegraphics[width=7cm]{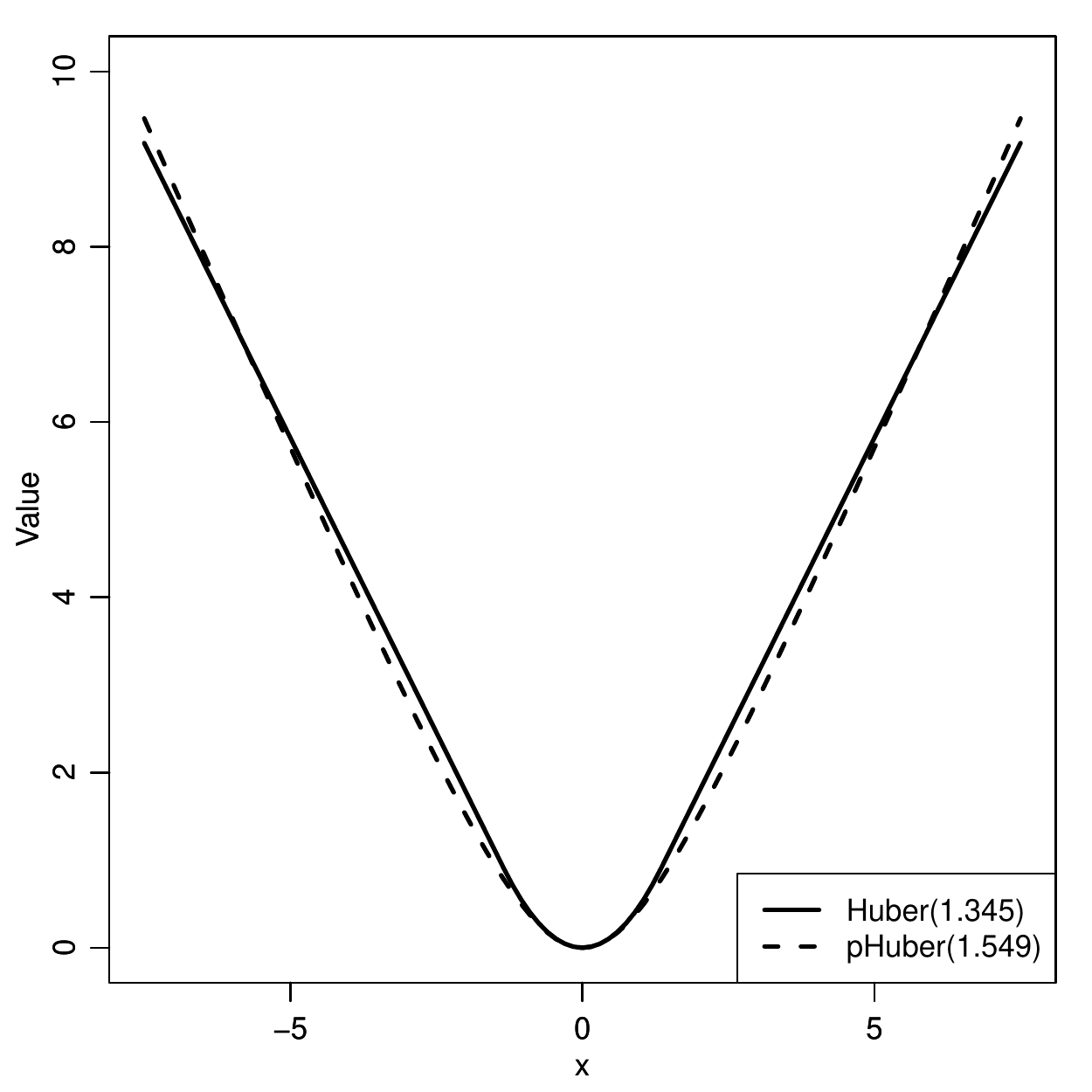}
\caption{Huber loss $L_c(x)$ and pseudo-Huber loss $L_c^{\mathrm{pH}}$ functions.}
\label{huber_loss_plot}
\end{figure}

The selection of tuning parameter for Huber-type loss functions is important and challenging issue. As pointed out by Peter Huber in his book (\cite{HR09}), ``The constant $c$ regulates the amount of robustness; good choices are in the range between $1$ and $2$, say, $c=1.5$". The default value of $c$ in {\tt R} package ({\tt rlm()} function in {\tt MASS} package) is $1.345$ which achieves about $95$\% asymptotic relative efficiency under the standard normal distribution. Although data-dependent procedures are also proposed by \cite{WLZB07} for example, the data-dependent selection of such tuning parameter is little known in the Bayesian community. In Figure \ref{huber_loss_plot}, we show the comparison of Huber loss with $c=1.345$ and pseudo-Huber loss $L_c^{\mathrm{pH}}(x)$ with $c\approx 1.549$ achieves about $95$\% asymptotic relative efficiency under the standard normal distribution.

\subsection{Bayesian Huberized lasso regression model}
\label{subsec:2.2}

In this paper, we consider a Bayesian analogue of Huberized lasso regression model. Following \cite{PC08}, the Bayesian Huberized lasso has the following hierarchical model by using the scale mixtures of normal representation for the hyperbolic distribution:
\begin{align}\label{model}
\begin{split}
\bm{y} \mid \mu,X,\bm{\beta},\bm{\sigma}^2 & \sim N_n(X\bm{\beta}, D_{\bm{\sigma}}),\\
\sigma_i^2 \mid \rho^2 & \overset{\mathrm{iid}}{\sim} \mathrm{GIG}(1,\eta,\rho^2)\quad (i=1,\dots,n; \  \eta>0),\\
\rho^2 & \sim \pi(\rho^2)\propto 1/\rho^2,\\
\bm{\beta} \mid \rho^2, \lambda &\sim \pi(\bm{\beta} \mid \rho^2) =\prod_{j=1}^p \frac{\lambda}{2\sqrt{\rho^2}}e^{-\lambda|\beta_j|/\sqrt{\rho^2}},
\end{split}
\end{align}
where $\bm{\sigma}^2:=(\sigma_1^2,\dots, \sigma_n^2)^{\top}$, $D_{\bm{\sigma}}:=\mathrm{diag}(\sigma_1^2,\dots,\sigma_n^2)$ and $\eta,\lambda>0$ are tuning parameters. The density function of $\mathrm{GIG}(1,\eta,\rho^2)$ is defined by \eqref{GIG2}. Following \cite{PC08}, we call the model \eqref{model} Bayesian Huberized lasso. As a prior of $\rho^2$, we assume the improper scale invariant prior which is proportional to $1/\rho^2$, but we can also employ a proper inverse gamma prior, for example. The posterior propriety for using the improper prior will be shown in Proposition \ref{propriety} later. By using the scale mixtures of normal representation of Laplace distribution (\cite{AM74}), we have
\begin{align}\label{Laplace}
\bm{\beta} \mid \bm{\tau}^2,\rho^2 \sim N_p(\bm{0},\rho^2 D_{\bm{\tau}}),\quad \tau_i^2 \mid \lambda \overset{\mathrm{iid}}{\sim} \mathrm{Exp}(\lambda^2/2) \quad (i=1,\dots,n),
\end{align}
where $\bm{\tau}^2:=(\tau_1^2,\dots,\tau_p^2)^{\top}$ and $D_{\bm{\tau}}:=\mathrm{diag}(\tau_1^2,\dots,\tau_p^2)$. 
As mentioned in \cite{PC08}, applying the model \eqref{model} and integrating out $\sigma_1^2,\dots,\sigma_n^2$ leads to the conditional density of $y$ given the remaining parameters as
\begin{align*}
\prod_{i=1}^n \frac{1}{2K_1(\eta)\sqrt{\eta \rho^2}} \exp\left(-\sqrt{\eta \left\{\eta+(y_i-\bm{x}_i^{\top}\bm{\beta})^2/\rho^2\right\}}\right)
\end{align*}
(see also \cite{G97}), which has the desired hyperbolic form. Here, we show two important properties of the model \eqref{model}. The proofs of Proposition \ref{propriety} and \ref{unimodality} are given in Appendix. 

\begin{prop}[Posterior propriety]\label{propriety}
Let $\rho^2 \sim \pi(\rho^2) \propto 1/\rho^2$ (improper scale-invariant prior). For fixed $\lambda>0$ and $\eta>0$, the posterior distribution is proper for all $n$.
\end{prop}

\begin{prop}[Posterior unimodality]\label{unimodality}
Under the conditional prior for $\bm{\beta}$ given $\rho^2$ and fixed $\lambda>0$ and $\eta>0$, the joint posterior $\pi(\bm{\beta},\rho^2 \mid \bm{y})$ is unimodal with respect to $(\bm{\beta},\rho^2)$. 
\end{prop}
Related to Proposition \ref{unimodality}, we will show that the unconditional prior for $\bm{\beta}$ can result in multimodality of the joint posterior in Subsection \ref{subsec:4.1}.

\section{Posterior computation via Markov chain Monte Carlo method}
\label{sec:3}

We construct an efficient posterior computation algorithm for the Bayesian Huberized lasso model. In Subsection \ref{subsec:3.1}, we introduce the standard Gibbs sampling algorithm when $\eta$ is fixed. After that, we propose a new approximate Gibbs sampling algorithm which also enables to estimate a tuning parameter $\eta$ from data. 

\subsection{Gibbs sampler for fixed $\eta$}
\label{subsec:3.1}

As mentioned by \cite{PC08}, the Gibbs sampler for the model \eqref{model} is easy to implement, but the detail  was not shown in their paper. In this subsection, we construct the Gibbs sampler for the model \eqref{model} for the sake of completeness. For the model \eqref{model} and the representation \eqref{Laplace}, the overall posterior distribution is given by 
\begin{align*}
&\pi(\bm{\beta},\bm{\tau}^2,\bm{\sigma}^2,\rho^2 \mid \bm{y})\\
&\propto \frac{1}{\sqrt{\sigma_1^2 \cdots \sigma_n^2}}\exp\left(-\frac{1}{2} (\bm{y}-X\bm{\beta})^{\top} D_{\bm{\sigma}}^{-1} (\bm{y}-X\bm{\beta})\right)  \times \prod_{i=1}^n \frac{1}{2\rho^2 K_1(\eta)} \exp\left(-\frac{\eta}{2}\left(\frac{\sigma_i^2}{\rho^2}+\frac{\rho^2}{\sigma_i^2}\right)\right) \\
&\quad \times \frac{1}{\sqrt{\rho^{2p} \tau_1^2\cdots \tau_p^2}} \exp\left(-\frac{1}{2\rho^2}\bm{\beta}^{\top}D_{\bm{\tau}}^{-1}\bm{\beta}\right)\times \prod_{j=1}^p \frac{\lambda^2}{2}e^{-\lambda^2 \tau_j^2/2} \times \frac{1}{\rho^2} 
\end{align*}
for $\eta>0$ and $\lambda>0$. Then the full conditional distributions are given by 
\begin{align}
\bm{\beta} \mid \bm{\tau}^2,\bm{\sigma}^2,\rho^2 & \sim N_p\left(A_{\bm{\tau},\bm{\sigma},\rho^2}^{-1}X^{\top}D_{\bm{\sigma}}^{-1}\bm{y},  A_{\bm{\tau},\bm{\sigma},\rho^2}^{-1}  \right),\label{full_cond_beta}\\
\rho^2 \mid \bm{\beta},\bm{\tau}^2,\bm{\sigma}^2& \sim \mathrm{GIG}\left(-n-p/2, \eta\sum_{i=1}^n \sigma_i^{-2}, \eta \sum_{i=1}^n \sigma_i^2 +\bm{\beta}^{\top}D_{\bm{\tau}}^{-1} \bm{\beta}\right),\label{full_cond_rho}\\
1/\tau_j^2 \mid \bm{\beta},\rho^2& \sim \mathrm{InvGauss}\left(\sqrt{\frac{\lambda^2 \rho^2}{\beta_j^2}},\lambda^2\right) \quad (j=1,\dots, p),\label{full_cond_tau}\\
1/\sigma_i^2 \mid \bm{\beta},\rho^2&\sim \mathrm{InvGauss}\left(\sqrt{\frac{\eta}{\rho^2\{(y_i-\bm{x}_i^{\top}\bm{\beta})^2 +\eta \rho^2\}}},\frac{\eta}{\rho^2}\right)  \quad (i=1,\dots,n)\label{full_cond_sigma}
\end{align}
for $\eta>0$ and $\lambda>0$, where $A_{\bm{\tau},\bm{\sigma},\rho^2}:=X^{\top}D_{\bm{\sigma}}^{-1}X+(1/\rho^2)D_{\bm{\tau}}^{-1}$. Note that the density function of the GIG in \eqref{full_cond_rho} is defined by \eqref{GIG1}. Hence, we can easily sample from the posterior $\pi(\bm{\beta},\bm{\tau}^2,\bm{\sigma}^2,\rho^2 \mid \bm{y})$ using the Gibbs sampler. When we assume the prior for $\lambda$, we may assume that $\lambda^2 \sim \mathrm{Ga}(a,b)$ ($a,b>0$), where $\mathrm{Ga}(a,b)$ denotes the gamma distribution with shape $a$ and rate $b$. As the same as \cite{PC08}, we note that the improper scale-invariant prior $1/\lambda^2$ for $\lambda^2$ leads to an improper posterior in the model \eqref{model}. Then we add the following full conditional distribution for $\lambda^2$ in the Gibbs sampler:
\begin{align}\label{full_cond_lam}
\lambda^2|\bm{\tau}^2 \sim \mathrm{Ga}\left(a+p, b+ \sum_{j=1}^p \tau_j^2/2\right).
\end{align}
The algorithm of Gibbs sampler which includes sampling for $\lambda^2$ are implemented through the following algorithm.

\subsubsection*{Gibbs sampler for fixed $\eta$}
Given the current state $\{(\bm{\beta}^{(s)}, \bm{\tau}^{2(s)}, \bm{\sigma}^{2(s)},\rho^{2(s)},\lambda^{2(s)})\}$, then we generate next state  as following:
\begin{enumerate}
\item Draw $\bm{\beta}^{(s+1)}$ from the full conditional distribution \eqref{full_cond_beta}. 
\item Draw $\rho^{2(s+1)}$ from the full conditional distribution \eqref{full_cond_rho}. 
\item Draw $(\bm{\tau}^{2(s+1)},\bm{\sigma}^{2(s+1)})$, independently.
\begin{itemize}
\item Draw $\bm{\tau}^{2(s+1)}$ from the full conditional distribution \eqref{full_cond_tau}.
\item Draw $\bm{\sigma}^{2(s+1)}$ from the full conditional distribution \eqref{full_cond_sigma}.
\end{itemize}
\item Draw $\lambda^{2(s+1)}$ from the full conditional distribution \eqref{full_cond_lam}.
\end{enumerate}
Note that we can also choose $\lambda$ using other methods. For example, \cite{PC08} also considered the empirical Bayes updating via Monte Carlo EM algorithm. Although to use the cross validation seems to be reasonable, we need to treat carefully to evaluate the prediction error in the presence of outliers. The selection of a tuning parameter $\lambda$ for the (non-robust) Bayesian lasso is also discussed in \cite{LN13}. Since we consider the fully Bayesian framework in this paper, we hereafter assume that $\lambda^2 \sim \mathrm{Ga}(a,b)$ for fixed $a$ and $b$. The remaining task is the selection of $\eta>0$ in the pseudo-Huber (or hyperbolic) loss function. We note that the parameter $\eta$ can also be thought as the prior hyper-parameter in model \eqref{model}.

\subsection{Proposal method: Approximate Gibbs sampler for estimation of $\eta$}
\label{subsec:3.2}

It is also important to choose a tuning parameter $\eta>0$ in practical use. In this paper, we consider the data-dependent selection of $\eta$ in fully Bayesian framework, that is, we assume a prior for $\eta$. However, the full conditional distribution of $\eta$ is intractable. In fact, if we assume a prior for $\eta$ as $\eta\sim \pi(\eta)$, then we have the full conditional distribution of $\eta$ as 
\begin{align}\label{fullcond_eta}
\pi(\eta\mid \bm{\sigma}^2,\rho^2)\propto \frac{1}{K_1(\eta)^n} \exp \left(-\frac{\eta}{2} \sum_{i=1}^n \left(\frac{\sigma_i^2}{\rho^2}+\frac{\rho^2}{\sigma_i^2}\right)\right)\times \pi(\eta).
\end{align}
Since the right-hand side of \eqref{fullcond_eta} involves the modified Bessel function of the second kind, the full conditional distribution of $\eta$ may not be a standard probability distribution for any prior $\pi(\eta)$. We now consider to approximate \eqref{fullcond_eta} by a standard probability distribution. 

We assume that $\eta$ has the gamma prior $\mathrm{Ga}(c,d)$ with shape $c$ and rate $d$, and consider the approximation of \eqref{fullcond_eta} by using the gamma distribution with shape $A$ and rate $B$ which is denoted by $\mathrm{Ga}(A,B)$. The idea comes from the paper by \cite{M19} which gives an approximation of the full conditional distribution for the shape parameter of the gamma distribution. The proposed algorithm is summarized in Algorithm \ref{algo:Gibbs2}. Although Steps 1 to 4 are the same as those of subsection \ref{subsec:3.1}, Step 5 is a new one to sample from the approximate full conditional distribution of $\eta$. The model considered here is summarized as follows:
\begin{align}\label{proposal}
\begin{split}
\bm{y} \mid \mu,X,\bm{\beta},\bm{\sigma}^2 & \sim N_n(X\bm{\beta}, D_{\bm{\sigma}}),\\
\sigma_i^2 \mid \rho^2 & \overset{\mathrm{iid}}{\sim} \mathrm{GIG}(1,\eta,\rho^2)\quad (i=1,\dots,n),\\
\rho^2 & \sim \pi(\rho^2)\propto 1/\rho^2,\\
\bm{\beta} \mid \bm{\tau}^2,\rho^2 &\sim N_p(\bm{0},\rho^2 D_{\bm{\tau}}),\\
\tau_i^2 \mid \lambda &\overset{\mathrm{iid}}{\sim} \mathrm{Exp}(\lambda^2/2) \quad (i=1,\dots,n),\\
\lambda^2 & \sim \mathrm{Ga}(a,b),\\
\eta &\sim \mathrm{Ga}(c,d),
\end{split}
\end{align}
where $a,b,c,d>0$ are fixed hyper-parameters. We use $a=b=c=d=1$ in simulation and real data analysis. The sensitivity analysis of the selection of hyper-parameters will also be discussed in Subsection \ref{subsec:4.4}. 

\begin{algorithm*}[t]

Set an iteration $M$ and a tolerance $\epsilon>0$ in Step 5. Given the current state, generate next state as following:

\begin{enumerate}
\item Draw $\bm{\beta}^{(s+1)}$ from the full conditional distribution \eqref{full_cond_beta}. 

\item Draw $\rho^{2(s+1)}$ from the full conditional distribution \eqref{full_cond_rho}.

\item Draw $(\bm{\tau}^{2(s+1)},\bm{\sigma}^{2(s+1)})$ independently.
\begin{itemize}

\item Draw $\bm{\tau}^{2(s+1)}$ from the full conditional distribution \eqref{full_cond_tau}.

\item Draw $\bm{\sigma}^{2(s+1)}$ from the full conditional distribution \eqref{full_cond_sigma}.

\end{itemize}

\item Draw $\lambda^{2(s+1)}$ from the full conditional distribution \eqref{full_cond_lam}.

\item Draw $\eta^{(s+1)}$ from $\mathrm{Ga}(A,B)$ given $(\bm{\sigma}^{2(s+1)},\rho^{2(s+1)})$, where $A$ and $B$ are given by the following algorithm:
\begin{enumerate}
\item Set the initial value as $A=a+n$ and $B=b+P$, where
\begin{align}\label{def_P}
P:=P(\bm{\sigma}^{2(s+1)},\rho^{2(s+1)})=\frac{1}{2}\sum_{i=1}^n\left(\frac{\sigma_{i}^{2(s+1)}}{\rho^{2(s+1)}}+\frac{\rho^{2(s+1)}}{\sigma_{i}^{2(s+1)}}\right).
\end{align}
\item For $k=1,\dots,M$ do
\begin{align}\label{AB}
\begin{split}
&\eta\leftarrow A/B\\
&A\leftarrow a+n\eta^2\frac{\partial^2}{\partial \eta^2}\log K_{1}(\eta)\\
&B \leftarrow b+\frac{A-a}{\eta}+n\frac{\partial}{\partial \eta}\log K_{1}(\eta)+P
\end{split}
\end{align}
If $|\eta/(A/B)-1|<\epsilon$, then return $A$ and $B$.
\end{enumerate}
\end{enumerate}
\caption{\bf--- Approximate Gibbs sampling for Bayesian Huberized lasso}
\label{algo:Gibbs2}
\end{algorithm*}

We now present the details of Step 5 in Algorithm \ref{algo:Gibbs2}. For simplicity, we reformulate the problem in this subsection. Assume that $x_1,\dots,x_n$ is a sequence of random variables according to the generalized inverse Gaussian distribution $\mathrm{GIG}(\nu,\eta,\rho^2)$ defined by \eqref{GIG2}, where $\nu$ and $\rho^2$ are known, and we assume the prior $\eta \sim \mathrm{Ga}(a,b)$. Let $\bm{x}=(x_1,\dots,x_n)$. Although the following result holds for $\nu \in \mathbb{R}$, we consider only $\nu=1$ for the application of the Bayesian Huberiozed lasso. So, we set $\nu=1$ for a while. We are interested in sampling from the posterior distribution of $\eta$, approximately. 

First of all, we explain the selection of an initial value in Step 5 (a). An initial values of $(A,B)$ are selected by approximating the modified Bessel function of the second kind. From \cite{AS65}, we have $K_{\nu}(x)\sim (1/2)\Gamma(\nu) (x/2)^{-\nu}$ as $x\to 0$ for $\nu>0$ and $K_{\nu}(x)\sim \sqrt{\pi/(2x)} e^{-x}$ as $x\to \infty$. Hereafter, we denote the true full conditional density of $\eta$ by $f(\eta)$. From the former approximation we have
\begin{align*}
f(\eta)=\pi(\eta\mid \bm{x},\rho^2=1)&\propto \frac{1}{K_1(\eta)^n} \exp \left(-\frac{\eta}{2} \sum_{i=1}^n \left(\frac{x_i}{\rho^2}+\frac{\rho^2}{x_i}\right)\right)\times \eta^{a-1}e^{-b\eta}\\
&\approx \eta^{a+n-1}e^{-(P+b)\eta},
\end{align*}
where $P$ is defined by \eqref{def_P}. Hence, it holds the approximation $f(\eta)\approx \mathrm{dgamma}(\eta\mid A=a+n,B=b+P)$ for small $\eta$, where $\mathrm{dgamma}(x\mid a,b)$ denotes the density function of a gamma distribution with shape $a$ and rate $b$. On the other hand, the latter one gives
\begin{align*}
f(\eta)=\pi(\eta\mid \bm{x},\rho^2=1)&\propto \frac{1}{K_1(\eta)^n} \exp \left(-\frac{\eta}{2} \sum_{i=1}^n \left(\frac{x_i}{\rho^2}+\frac{\rho^2}{x_i}\right)\right)\times \eta^{a-1}e^{-b\eta}\\
&\propto \left(\frac{\pi}{2}\right)^{-n/2}\eta^{n/2}e^{n\eta}\exp \left(-\frac{\eta}{2} \sum_{i=1}^n \left(\frac{x_i}{\rho^2}+\frac{\rho^2}{x_i}\right)\right)\times \eta^{a-1}e^{-b\eta}\\
&\approx \eta^{a+n/2-1}e^{-(P+b-n)\eta}.
\end{align*}
Therefore, it hold the approximation $f(\eta)\approx \mathrm{dgamma}\left(\eta\mid A=a+(n/2),B=b+P-n\right)$ for large $\eta$. 
We adopt the former one ($A=a+n$ and $B=b+P$) as an initial value for Step 5 in Algorithm \ref{algo:Gibbs2}. If we use the later one ($A=a+(n/2)$ and $B=b+P-n$) as an initial value of algorithm, the result did not change very much. In practical use, we may choose $M=10$ and a tolerance $\epsilon=10^{-8}$. The updating step \eqref{AB} in Step 5 (b) is given by the similar method as \cite{M19} which matches the first- and second-derivatives of logarithm of the true full conditional density $\log f(\eta)$ and that of the gamma density denoted by $\log g(\eta)$ with shape $A$ and rate $B$. In our case, we have
\begin{itemize}
\item $f(\eta)=\pi(\eta\mid x_{1},\dots,x_{n},\rho^2=1)$
\begin{align*}
\log f(\eta)&=\text{const.}+\log f(\bm{x}\mid\eta)+\log\pi(\eta)\\
&=\text{const.}-n\log K_1(\eta)-\eta P+(a-1)\log \eta-b\eta\\
\frac{\partial}{\partial \eta}\log f(\eta)&=-n\frac{\partial}{\partial \eta}\log K_{1}(\eta)+\frac{a-1}{\eta}-P-b\\
\frac{\partial^2}{\partial \eta^2}\log f(\eta)&=-n\frac{\partial^2}{\partial \eta^2}\log K_{1}(\eta)-\frac{a-1}{\eta^2}
\end{align*}
\item $g(\eta)=\text{dgamma}(\eta\mid\text{shape}=A,\text{scale}=1/B)$
\begin{align*}
\log g(\eta)&=A\log B-\log \Gamma(A)+(A-1)\log \eta-B\eta\\
\frac{\partial}{\partial \eta}\log g(\eta)&=\frac{A-1}{\eta}-B\\
\frac{\partial^2}{\partial \eta^2}\log g(\eta)&=-\frac{A-1}{\eta^2}
\end{align*}
\end{itemize}
Hence, we have the following equations:
\begin{align*}
&\frac{\partial^2}{\partial \eta^2}\log f(\eta)=\frac{\partial^2}{\partial \eta^2}\log g(\eta)\iff -n\frac{\partial^2}{\partial \eta^2}\log K_{1}(\eta)-\frac{a-1}{\eta^2}=-\frac{A-1}{\eta^2},\\
&\frac{\partial}{\partial \eta}\log f(\eta)=\frac{\partial}{\partial \eta}\log g(\eta)\iff -n\frac{\partial}{\partial \eta}\log K_{1}(\eta)+\frac{a-1}{\eta}-P-b=\frac{A-1}{\eta}-B.
\end{align*}
By solving equations with respect to $A$ and $B$, we have
\begin{align*}
A=a+n\eta^2\frac{\partial^2}{\partial \eta^2}\log K_{1}(\eta),\quad B=b+\frac{A-a}{\eta}+n\frac{\partial}{\partial \eta}\log K_{1}(\eta)+P.
\end{align*}
This algorithm is supported by characterizing its fixed points.
\begin{prop}\label{fixed_point}
For any $\eta>0$, $\eta$ is a fixed point for \eqref{AB} in Algorithm \ref{algo:Gibbs2} if and only if it holds that
\begin{align}\label{identity_fixed_point}
\frac{\partial}{\partial \eta}\log f(\eta)+\frac{1}{\eta}=0.
\end{align}
Furthermore, there exists $\eta>0$ such that $(\partial/\partial \eta)\log f(\eta)+(1/\eta)=0$.
\end{prop}

Details of the interpretation of this result are mentioned by the Supplementary material of \cite{M19}. The identity has a natural interpretation in terms of optimization with a logarithmic barrier. Although the proof of Proposition \ref{fixed_point} is similar to that of \cite{M19}, the proofs of the technical parts are new, and our result is a kind of generalizations of the result in \cite{M19}. In other words, although they considered only the case of the approximation of the full conditional distribution for a shape parameter of the gamma distribution, our result shows that the result of \cite{M19} can be extended to the full conditional distribution for a concentration parameter $\eta$ of generalized inverse gaussian distribution $\mathrm{GIG}(\nu,\eta,\rho^2)$ for any $\nu \in \mathbb{R}$ and $\rho^2>0$ which includes gamma distribution as a special case. We give the proof in Appendix. 

As another selection method for $\eta$, we might use the random-walk Metropolis-Hasting (RWM) algorithm. Since $\eta$ is one-dimension, the tuning for the proposal distribution is not difficult while the ratio of true full conditionals is numerically unstable because of the need of evaluation for the modified Bessel function of the second kind. Hence, we do not discuss the RWM algorithm here.

\begin{figure}[htpb]
\centering
\includegraphics[width=15cm]{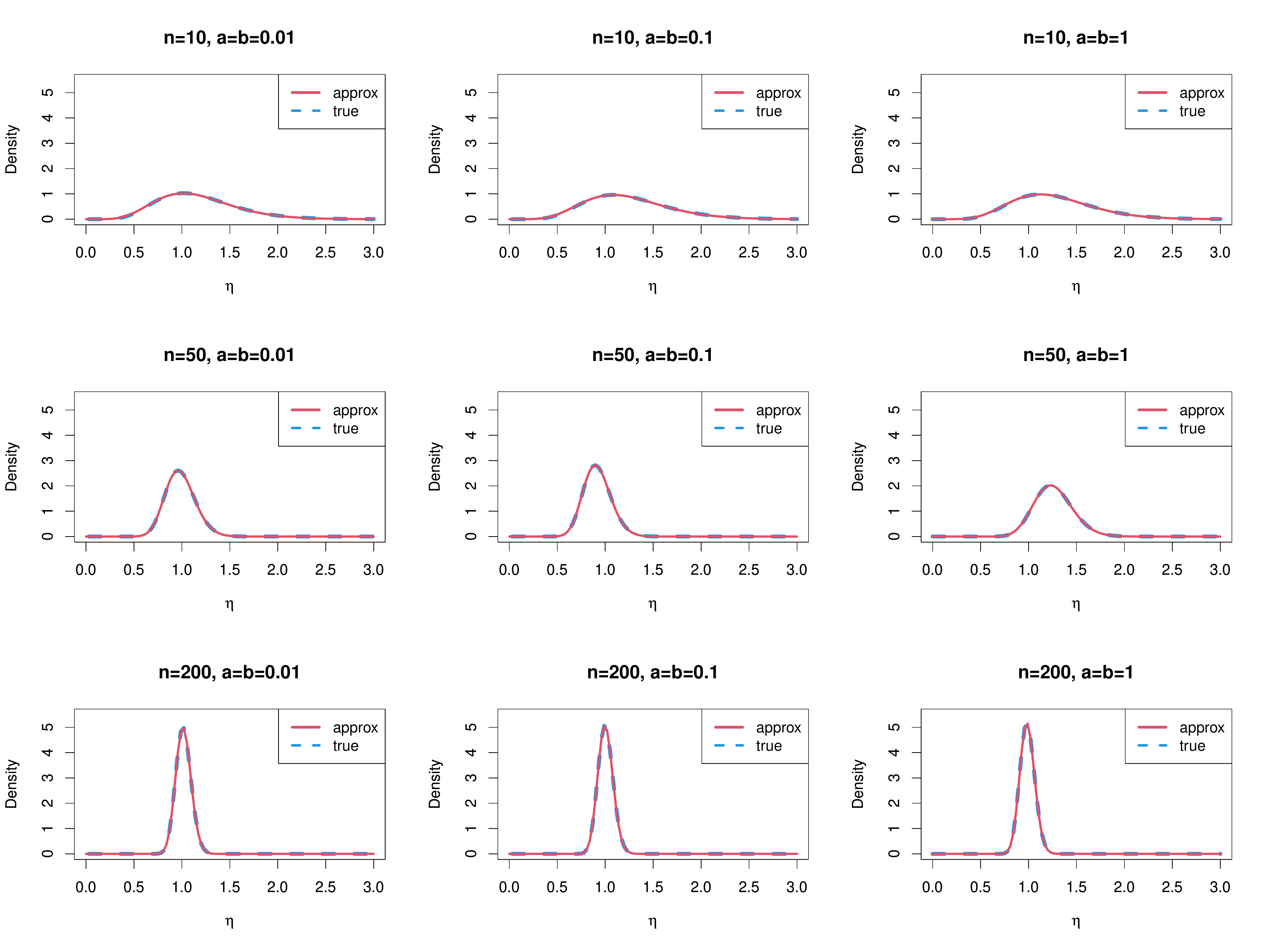}
\caption{Probability density function (PDF) of the true full conditional and the approximate full conditional on simulated data $x_1,\dots,x_n \sim \mathrm{GIG}(1,\eta,1)$ when the prior is $\eta\sim \mathrm{Ga}(a,b)$, where the true value of $\eta$ is $1$.}
\label{density_plot}
\end{figure}

\begin{figure}[htbp]
\centering
\includegraphics[width=15cm]{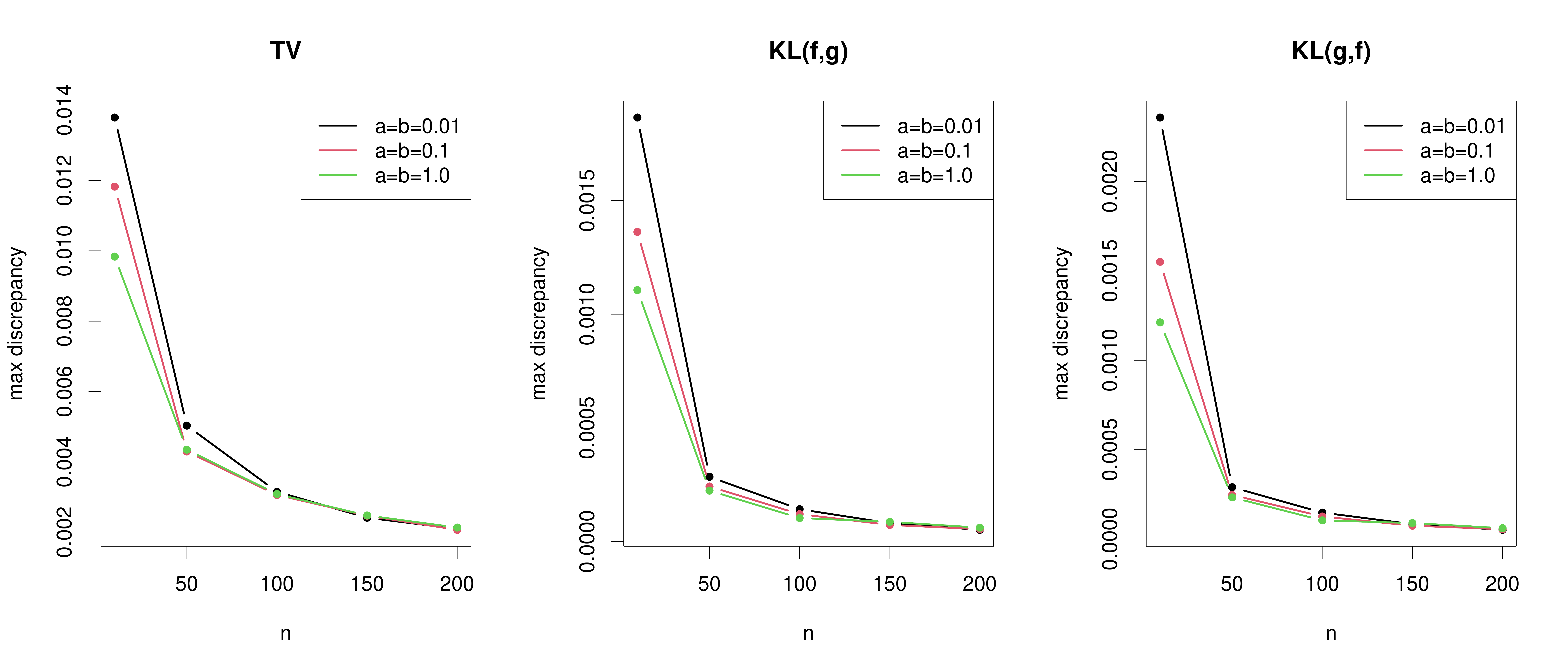}
\caption{Largest observed discrepancy between the true full conditional $f$ and the approximate full conditional $g$ for the total variation distance (TV), the Kullback-Leibler divergence ($\mathrm{KL}(f,g)$), and  the reverse Kullback-Leibler divergence ($\mathrm{KL}(g,f)$).}
\label{distance_plot}
\end{figure}

We now check the approximation accuracy of the full conditional distribution for a concentration parameter of the generalized inverse gaussian distribution through some numerical studies. We are interested in the approximation full conditional distribution of $\eta$ by using the gamma distribution for 
\[\sigma_i^2 \mid \rho^2  \overset{\mathrm{iid}}{\sim} \mathrm{GIG}(1,\eta,\rho^2)\quad (i=1,\dots,n; \  \eta>0),\quad \eta\sim \mathrm{Ga}(c,d)\]
in the model \eqref{proposal}. We now use more simple notations as follows:
\[x_1,\dots,x_n \overset{\mathrm{iid}}{\sim} \mathrm{GIG}(1,\eta,1),\quad \eta\sim \mathrm{Ga}(a,b).\]
Following \cite{M19}, we set $a=b$ in this simulation and we assume that simulation dataset is generated from $\mathrm{GIG}(\nu=1,\eta=1,\rho^2=1)$. In this case, the true full conditional distribution $f$ is given by
\begin{align*}
f(\eta)\propto \frac{1}{K_1(\eta)^n} \exp \left(-\frac{\eta}{2}\sum_{i=1}^n \left(x_i+\frac{1}{x_i}\right)\right)\times \eta^{a-1}e^{-b\eta}
\end{align*}
and the approximate full conditional distribution $g$ is $\mathrm{Ga}(A,B)$ distribution, where $A$ and $B$ are given by Algorithm \eqref{AB}. Note that using other values of $(\nu,\eta,\rho^2)$ as data generation distribution did not change very much in the following simulation results. Figure \ref{density_plot}, we show the probability density functions the true full conditional and the approximate full conditional for $n \in \{10,50,200\}$ and $a=b\in \{0.01,0.1,1\}$. In each case, the approximation is close to the true full conditional distribution. To quantify the discrepancy between the true and approximate full conditionals, we use three famous discrepancy measures, the total variation distance $d_{\mathrm{TV}}(f,g)=(1/2)\int |f(\eta) -g(\eta)| d\eta$, the Kullback-Leibler divergence $d_{\mathrm{KL}}(f,g)=\int f(\eta)\log(f(\eta)/g(\eta)) d\eta$ and the reverse Kullback-Leibler divergence $d_{\mathrm{KL}}(g,f)=\int g(\eta)\log(g(\eta)/f(\eta))d\eta$. For $n \in \{10,50,100,150,200\}$ and $a=b\in \{0.01,0.1,1\}$, Figure \ref{distance_plot} shows the maximum discrepancy over one hundred dataset for each case. In all case, the approximation accuracy improves for large $n$, and the accuracy also improves for large $a$ and $b$. Calculations in Figures \ref{density_plot} and \ref{distance_plot} is based on importance sampling which is the same as the Section S3 in Supplementary material of \cite{M19}.

\section{Simulation studies}
\label{sec:4}

In this section, we give some simulation studies on the proposed method.

\subsection{Multimodality of joint posterior}
\label{subsec:4.1}

\begin{figure}[htpb]
\centering
\includegraphics[width=14cm]{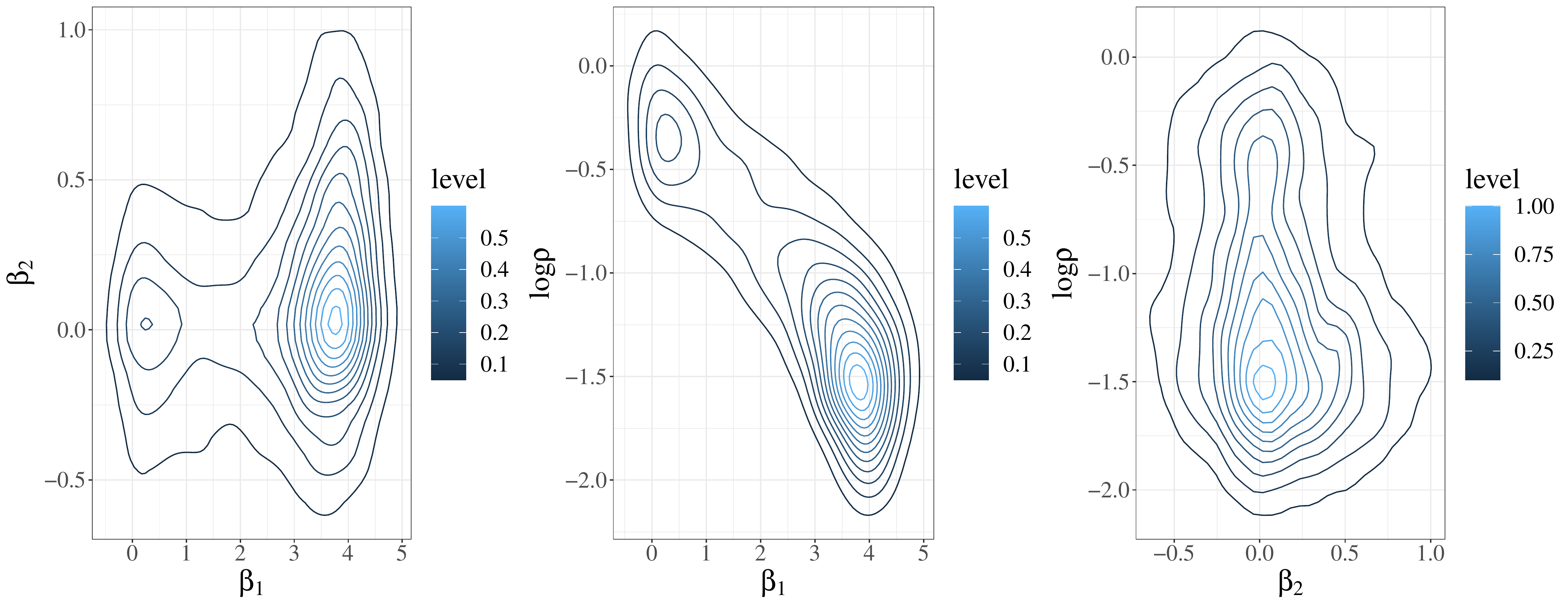}
\caption{Pairwise contour plot of posterior densities with $\lambda=3$ and $\eta=1$.} 
\label{fig:multimodal}
\end{figure}

As related to Proposition \ref{unimodality}, we present a simple simulation to demonstrate that the unconditional prior for $\bm{\beta}$ can result in multimodality of the joint posterior. The unconditional prior is defined by
\[ \pi(\bm{\beta}) =\prod_{j=1}^p \frac{\lambda}{2}e^{-\lambda|\beta_j|}\]
instead of the prior for $\bm{\beta}$ in the model \eqref{model}, and we now assume that $\lambda>0$ and $\eta>0$ are fixed. The corresponding Gibbs sampler is easily obtained. 
We generate the data with the linear model as $\bm{y}=X\bm{\beta}+\sigma \varepsilon$, where $\epsilon_1,\dots,\epsilon_n \overset{\mathrm{iid}}{\sim} \mathrm{Hyp}(0 , 1 , 1 , 0)$ and $\sigma=0.03$. In a similar way to \cite{CS21}, we take $\bm{\beta}=(0,5)$, $\mathrm{tr}(X^{\top}X)=1$, $\lambda=3$, and $\eta=1$. In Figure \ref{fig:multimodal}, we can see that joint posterior densities of $(\beta_1,\beta_2)$, $(\beta_1,\log \rho)$ and $(\beta_2, \log \rho)$ are multi-modal. If the posterior is multi-modal, it may slow down the convergence of the Gibbs sampler and the point estimate may be computed through multiple modes, which leads to the inaccurate estimators (see also \cite{PC08}, \cite{KGGC10} and \cite{CS21}). For these reasons, we use the prior for $\bm{\beta}$ conditioned on scale parameter in this article. 

\subsection{Bayesian robustness properties using influence function}
\label{subsec:4.2}

\begin{figure}[htpb]
\centering
\includegraphics[width=13cm]{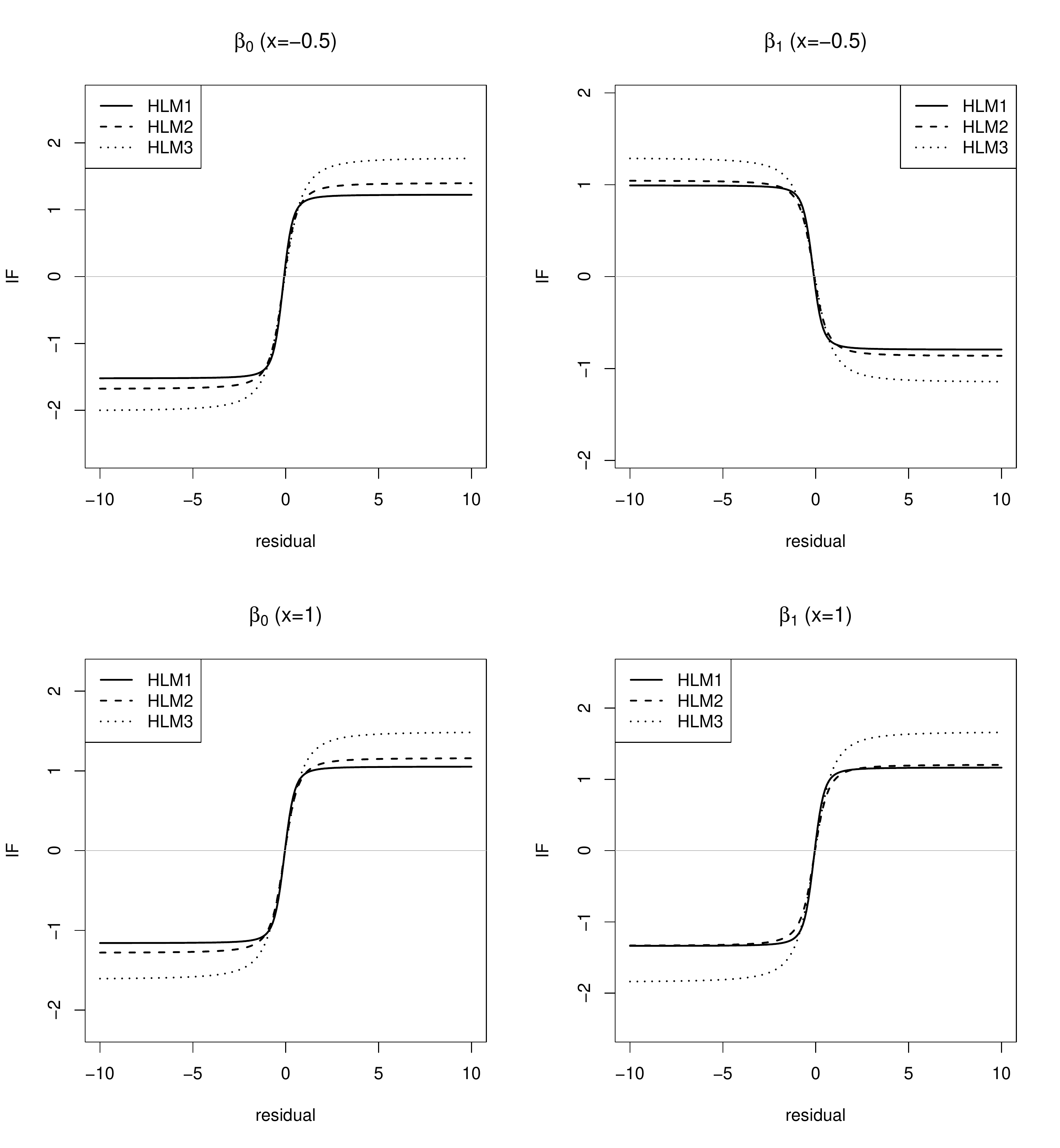}
\caption{Influence functions for $\beta_0$ (intercept) and $\beta_1$ (regression coefficient) in a simple Bayesian linear regression model under uniform prior for $(\beta_0,\beta_1)$ (HLM: Bayesian Huberized linear regression model for $\eta=0.2$ (HLM1), $\eta=0.5$ (HLM2), and $\eta=1.0$ (HLM3). }
\label{IF_plot}
\end{figure}

We check robustness of the posterior mean under the Bayesian Huberized linear (denoted by HLM here) regression model through simple simulation studies. As a measure of robustness, we use the Bayesian influence function considered in \cite{GB16} and \cite{NH20}. Following \cite{HS20}, we consider a simple linear regression model $y_i=\beta_0+\beta_1 x_i+\varepsilon_i$ ($i=1,\dots,n$), where $\beta_0=0$, $\beta_1=1$, $\varepsilon_i\sim N(0,1)$, $x_i$ is generated from the standard normal distribution, and we set $n=100$. We used uniform priors for $\beta_0$ and $\beta_1$ to obtain the posterior distribution of $\theta=(\beta_0,\beta_1)^{\top}$ under the model $y_i=\beta_0+\beta_1 x_i+\varepsilon_i$. The influence function of the posterior means of $\beta_k$ ($k=0,1$) is defined by $\mathrm{IF}_k(z\mid x)=n\mathrm{Cov}_{\theta\mid D}(\beta_k, H(\theta,z\mid x))$, where $\mathrm{Cov}_{\theta\mid D}$ denotes the covariance with respect to the posterior of $\theta$ given data $D$ (see \cite{GB16}, \cite{NH20} and \cite{HS20}).  
Under the standard likelihood function, it holds that $H(\theta, z \mid x)=\log f(\beta_0+\beta_1 x+z \mid x;\theta)-\int \log f(t \mid x;\theta) g(t \mid x)dt$, where $g(\cdot \mid x)=\phi(\cdot; x, 1)$ is the true density. We use a hyperbolic distribution as $f(\beta_0+\beta_1 x+z|x;\theta)$. We note that $z$ can be interpreted as the residual of the outlying value, namely, the distance between the outlying value and the regression line $\beta_0+\beta_1 x$.
We approximated the integral appeared in $H$ by Monte Carlo integration based on 2000 random samples from $g(\cdot \mid x)$. Based on 10000 posterior samples of $\theta$, we computed ${\rm IF}_0(z \mid x)$ and ${\rm IF}_1(z \mid x)$ which are the influence functions for $\beta_0$ and $\beta_1$, respectively, for $x\in \{-0.5, 1\}$ and $z\in [-10, 10]$, under the hyperbolic loss with $\eta=0.2$ (HLM1), $\eta=0.5$ (HLM2) and $\eta=1.0$ (HLM3). The results are presented in Figure \ref{IF_plot}. The posterior means of Bayesian Huberized linear regression models have bounded influence functions. From Figure \ref{IF_plot}, we can see that the smaller $\eta$, the smaller influence from an outlier.

\subsection{Simulation studies}
\label{subsec:4.3}

In simulation studies, we illustrate performance of the proposed method. 
We compare the point and interval estimation performance of the proposed method with those of some existing methods.
To this end, we consider the following regression model with $n \in \{100, 150, 200\}$ and $p=20$:
\begin{align*}
y_i=\beta_0 + \beta_1x_{i1}+\cdots+\beta_px_{ip}+\sigma \varepsilon_i  \quad  (i=1,\ldots,n),
\end{align*}
where $\beta_0=1$, $\beta_1=3$,  $\beta_2=0.5$, $\beta_4=\beta_{11}=1$, $\beta_7=1.5$ and the other $\beta_j$'s were set to $0$. We assume that $\bm{y}=(y_1,\dots,y_n)^{\top}$ is the response vector. 
The covariates $\bm{x}_i=(x_{i1},\ldots,x_{ip})$ were generated from a multivariate normal distribution $N_p(0,\Sigma)$ with $\Sigma=(r^{|i-j|})_{1\leq i,j\leq p}$ for $|r|<1$. Following \cite{LZ11}, we consider the four scenarios.
\begin{itemize}
\item Model 1: Low correlation and Gaussian noise. $\bm{\varepsilon}\sim N_n(0,I_n)$, $\sigma=2$ and $r=0.5$.
\item Model 2: High correlation and Gaussian noise. $\bm{\varepsilon}\sim N_n(0,I_n)$, $\sigma=2$ and $r=0.95$.
\item Model 3: Large outliers. $\varepsilon=V/\sqrt{\mathrm{var}(V)}$, $\sigma=9.67$ and $r=0.5$. $V$ is a random variable according to the contaminated density defined by $0.9 \times N(0,1) + 0.1 \times N(0,225)$, where $\sqrt{\mathrm{var}(V)}=4.83$. 
\item Model 4: Sensible outliers. $\varepsilon=D/\sqrt{\mathrm{var}(D)}$, $\sigma=9.67$ and $r=0.5$. $D$ is a random variable according to the Laplace density, that is, $D\sim \mathrm{Laplace}(0,1)=e^{-|x|}/2$, where $\mathrm{var}(D)=2$.
\end{itemize}

For the simulated dataset, we applied the proposed robust method denoted by HBL as well as the standard (non-robust) Bayesian lasso (BL) by \cite{PC08}.
Moreover, as existing robust methods, we also applied the regression model with the error term following $t$-distribution with $3$ degrees of freedom and Bayesian median regression, denoted by tBL and mBL, respectively. The mBL is a Bayesian alternative to the LAD regression. For the mBL, we obtain posterior samples using the Gibbs sampling for the Bayesian quantile regression, where the quantile level is equal to $1/2$ (see also \cite{KK11}). For the tBL, the Gibbs sampling is easily obtained by the scale mixtures representation of $t$-distribution. In BL, tBL and mBL, we assume that $\sigma^2 \sim p(\sigma^2) \propto 1/\sigma^2$ (uninformative prior) and $\lambda^2 \sim \mathrm{Ga}(a=1,b=1)$. For the HBL, we calculate the posterior distribution using Algorithm \ref{algo:Gibbs2} in Subsection \ref{subsec:3.2}, where iteration of Step 5 is $M=10$, a tolerance is $\epsilon=10^{-8}$, and the prior is $\eta\sim \mathrm{Ga}(c=1,d=1)$. The sensitivity analysis of hyper-parameters will be  given in Subsection \ref{subsec:4.4} later. 
In applying the above methods, we generated 2000 posterior samples after discarding the first 500 samples as burn-in.
For point estimates of $\beta_j$'s, we computed posterior median of each element of $\beta_j$'s, and the performance is evaluated via root of mean squared error (RMSE) defined as $\{(p+1)^{-1}\sum_{j=0}^p(\widehat{\beta}_j-\beta_j)^2\}^{1/2}$.  
We also computed $95\%$ credible intervals of $\beta_j$'s, and calculated average lengths (AL) and coverage probability (CP) defined as $(p+1)^{-1}\sum_{j=0}^p|{\rm CI}_j|$ and $(p+1)^{-1}\sum_{j=0}^p I(\beta_j\in {\rm CI}_j)$, respectively. These values were averaged over 300 replications of simulating datasets.

Since the purpose of this paper is to  propose robust and efficient shrinkage estimation of regression coefficients, we do not discuss details of variable selection. Although one approach could be to set coefficients to zero when zero lies within the 95\% credible interval (e.g. \cite{PC08}, \cite{FKK10}), this approach does not take account of model uncertainly and it depends on the construction of the credible intervals. The Bayesian variable selection methods for sparse Bayesian linear regression models have been also developed by many authors (see e.g. \cite{GM97}, \cite{KM98}, \cite{LN13}).

\begin{figure}[htpb]
\centering
\includegraphics[width=13cm]{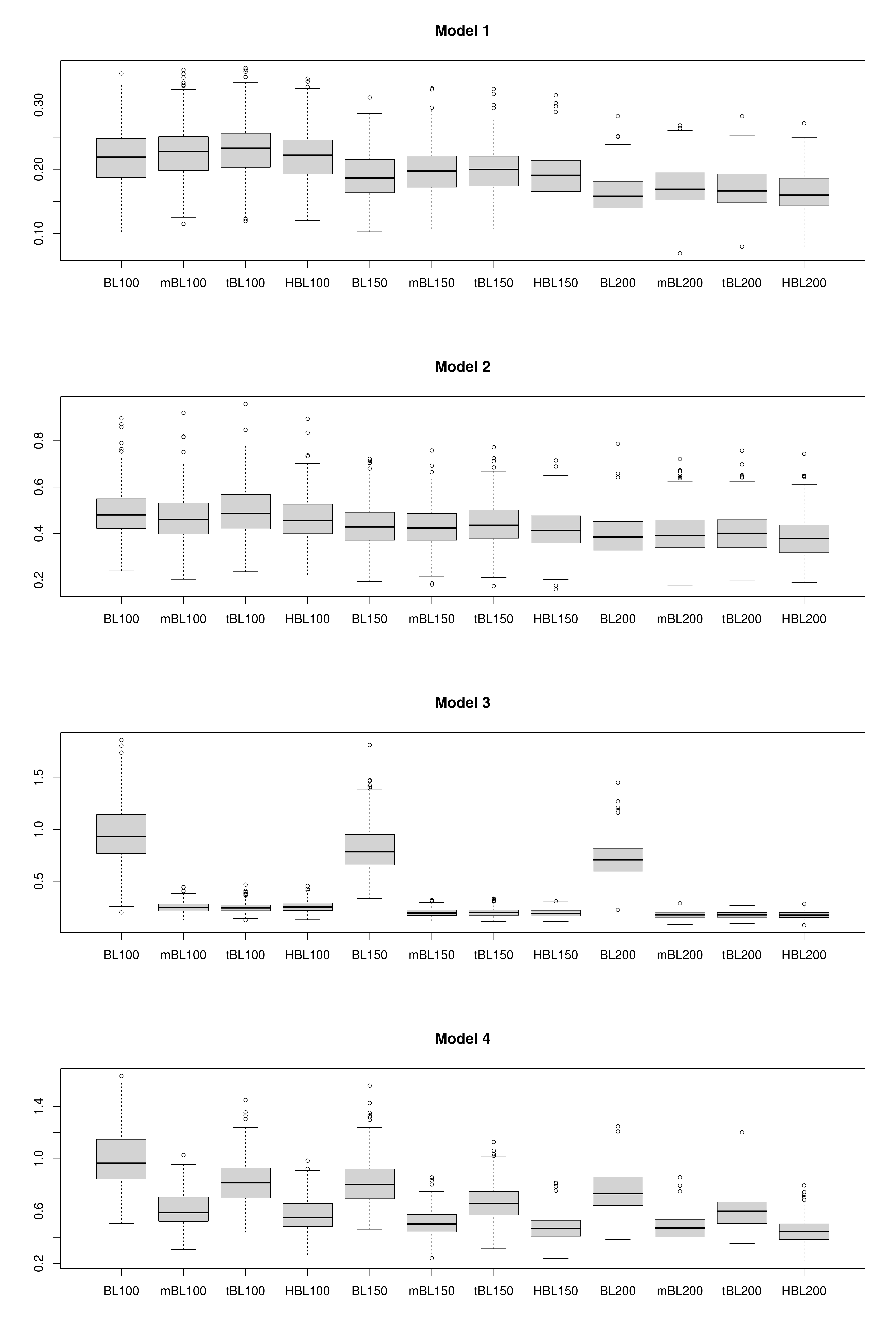}
\caption{Boxplot of RMSE based on 300 replications in four simulation scenarios}
\label{fig:RMSE}
\end{figure}

\begin{figure}[htpb]
\centering
\includegraphics[width=13cm]{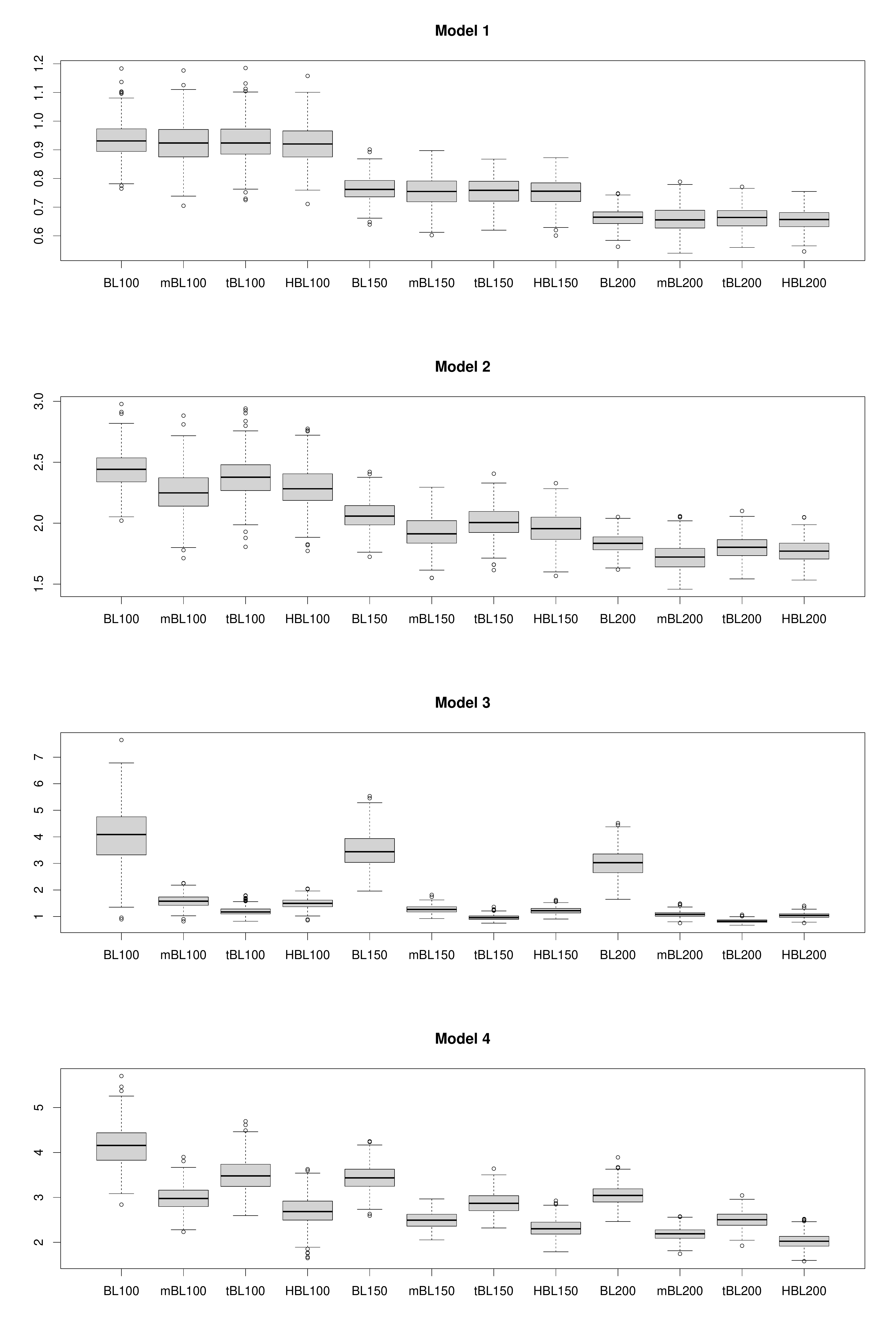}
\caption{Boxplot of AL based on 300 replications in four simulation scenarios}
\label{fig:AL}
\end{figure}

\begin{figure}[htpb]
\centering
\includegraphics[width=10cm]{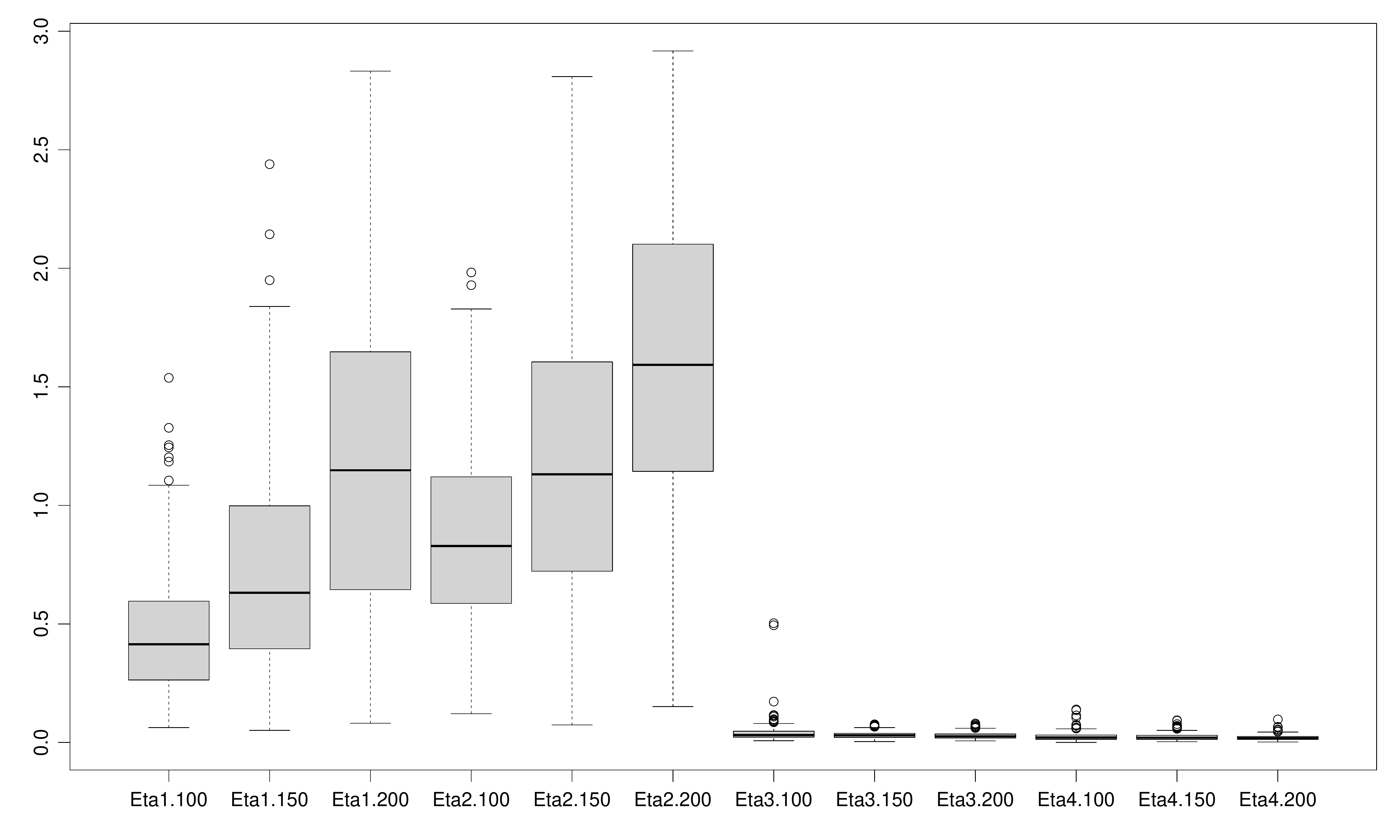}
\caption{Boxplot of posterior median of $\eta$ in the proposed method (HBL) based on 300 replications in four simulation scenarios}
\label{fig:eta}
\end{figure}

We report simulation results in Tables \ref{tab:model1} to \ref{tab:model4} and Figures  \ref{fig:RMSE} and \ref{fig:AL}. In Model 1 and Model 2, there is no outliers in simulated datasets. In such cases, the original Bayesian lasso works well, and has the smallest RMSE  for large $n$. On the other hand, since the RMSEs of HBL are smaller than those of mBL and tBL, we can find that HBL is more efficient than mBL and tBL. Such results are reasonable because the HBL involved the BL as the special case as $\eta\to \infty$. In the presence of large outliers (Model 3), tBL is better than other methods because the influence function of the estimator under t-error has the redescending property (see \cite{HRRS11}). In this simulation, we used the $t$-distribution with degree of freedom $3$. However, we note that the selection of degree of freedom for $t$-distribution needs to be set carefully in practical use. On the other hand, mBL and HBL are comparable in Model 3. For sensible outliers (Model 4), while the HBL is better than other competitors in terms of RMSE, tBL is slightly worse than mBL and HBL. The CPs are also reasonable in Model 1,2, and 4, but it seems to be over coverage in Model 3. We also show the boxplot of posterior median of $\eta$ in Figure \ref{fig:eta}. In the absence of outliers (Model 1 and 2), the posterior median of $\eta$ has large values. In the presence of large outliers (Model 3 and 4), small $\eta$ is selected. Hence, we can see that $\eta$ is adaptively selected for each dataset. In summary, although the proposed HBL seems to be conservative, it usually performs well in all scenarios. 

\begin{table}[h]
  \begin{minipage}[t]{.45\textwidth}
  \caption{Numerical results in Model 1}
    \begin{center}
      \begin{tabular}{rrrr}
  \hline
 & RMSE  & AL & CP  \\ 
  \hline
BL100 & {\bf 0.220}  & 0.935 & 0.962  \\ 
  mBL100 & 0.227  & 0.926 & 0.955  \\ 
  tBL100 & 0.232  & 0.928 & 0.950  \\ 
  HBL100 & 0.221  & 0.921 & 0.959  \\ \hline
  BL150 & {\bf 0.189}  & 0.764 & 0.949  \\ 
  mBL150 & 0.197  & 0.754 & 0.940  \\ 
  tBL150 & 0.199  & 0.758 & 0.940  \\ 
  HBL150 & 0.191  & 0.754 & 0.949  \\ \hline
  BL200 & {\bf 0.162}  & 0.664 & 0.954  \\ 
  mBL200 & 0.174  & 0.657 & 0.934  \\ 
  tBL200 & 0.172  & 0.662 & 0.938  \\ 
  HBL200 & 0.165  & 0.657 & 0.947  \\ 
   \hline
      \end{tabular}
    \end{center}
    \label{tab:model1}
  \end{minipage}
  \hfill
  \begin{minipage}[t]{.45\textwidth}
  \caption{Numerical results in Model 2}
    \begin{center}
      \begin{tabular}{rrrr}
  \hline
 & RMSE  & AL & CP  \\ 
  \hline
BL100 & 0.489  & 2.440 & 0.981  \\ 
  mBL100 & 0.469  & 2.258 & 0.976  \\ 
  tBL100 & 0.498  & 2.377 & 0.973  \\ 
  HBL100 & {\bf 0.462}  & 2.295 & 0.979  \\ \hline
  BL150 & 0.434  & 2.062 & 0.979  \\ 
  mBL150 & 0.427  & 1.921 & 0.971  \\ 
  tBL150 & 0.442  & 2.011 & 0.974  \\ 
  HBL150 & {\bf 0.418}  & 1.961 & 0.978  \\ \hline
  BL200 & 0.395  & 1.837 & 0.971  \\ 
  mBL200 & 0.403  & 1.725 & 0.957  \\ 
  tBL200 & 0.409  & 1.803 & 0.966  \\ 
  HBL200 & {\bf 0.387}  & 1.772 & 0.970  \\ 
   \hline      
      \end{tabular}
    \end{center}
    \label{tab:model2}
  \end{minipage}
\end{table}

\begin{table}[h]
  \begin{minipage}[t]{.45\textwidth}
  \caption{Numerical results in Model 3}
    \begin{center}
      \begin{tabular}{rrrr}
  \hline
 & RMSE  & AL & CP  \\ 
  \hline
BL100 & 0.965  & 4.101 & 0.959  \\ 
  mBL100 & 0.252  & 1.581 & 0.995  \\ 
  tBL100 & {\bf 0.250}  & 1.198 & 0.979  \\ 
  HBL100 & 0.255  & 1.495 & 0.995  \\ \hline
  BL150 & 0.820  & 3.495 & 0.958  \\ 
  mBL150 & 0.198  & 1.273 & 0.996  \\ 
  tBL150 & 0.199  & 0.965 & 0.978  \\ 
  HBL150 & {\bf 0.195}  & 1.218 & 0.995  \\ \hline
  BL200 & 0.722  & 3.014 & 0.955  \\ 
  mBL200 & 0.178  & 1.084 & 0.996  \\ 
  tBL200 & 0.176  & 0.827 & 0.977  \\ 
  HBL200 &{\bf  0.174}  & 1.041 & 0.995  \\ 
   \hline
      \end{tabular}
    \end{center}
    \label{tab:model3}
  \end{minipage}
  \hfill
  \begin{minipage}[t]{.45\textwidth}
  \caption{Numerical results in Model 4}
    \begin{center}
      \begin{tabular}{rrrr}
  \hline
 & RMSE  & AL & CP  \\ 
  \hline
BL100 & 1.001  & 4.161 & 0.957  \\ 
  mBL100 & 0.610  & 2.990 & 0.977  \\ 
  tBL100 & 0.825 & 3.491 & 0.958  \\ 
  HBL100 & {\bf 0.575}  & 2.707 & 0.972  \\ \hline
  BL150 & 0.819 & 3.449 & 0.961  \\ 
  mBL150 & 0.512  & 2.496 & 0.977  \\ 
  tBL150 & 0.670  & 2.870 & 0.959  \\ 
  HBL150 & {\bf 0.478}  & 2.313 & 0.974  \\ \hline
  BL200 & 0.754  & 3.051 & 0.951  \\ 
  mBL200 & 0.477  & 2.184 & 0.968  \\ 
  tBL200 & 0.599  & 2.507 & 0.956  \\ 
  HBL200 & {\bf 0.449}  & 2.026 & 0.967  \\ 
   \hline      
      \end{tabular}
    \end{center}
    \label{tab:model4}
  \end{minipage}
\end{table}

\begin{figure}[htpb]
\centering
\includegraphics[width=7cm]{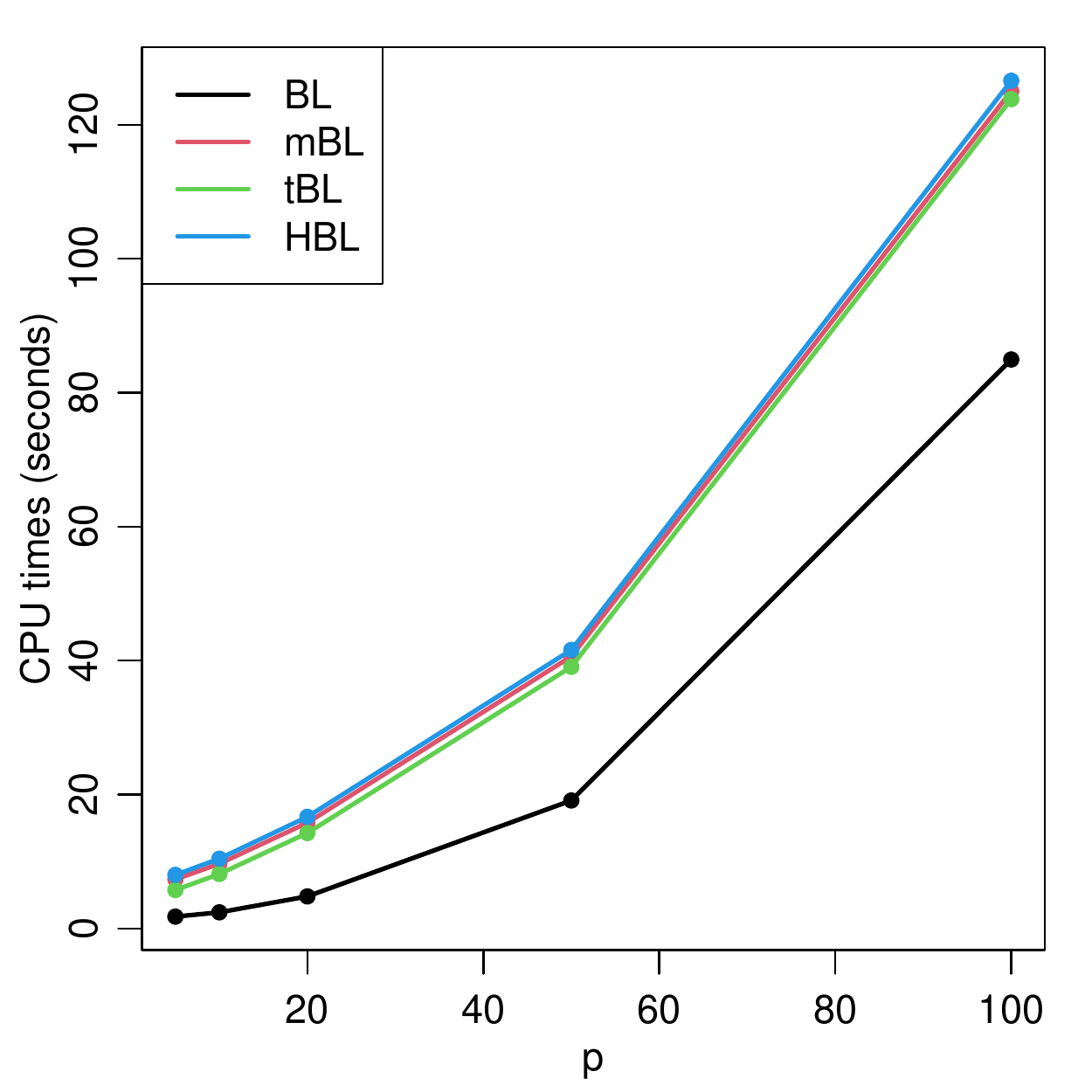}
\caption{CPU times (in seconds) for $n = 200$ and varying $p\in \{5,10,20,50,100\}$, averaged over $10$ runs which are generated 10000 posterior samples for each steps.}
\label{fig:CPU}
\end{figure}

\begin{rem}[Computational details and CPU times]
We compare the computation times for MCMC based on BL, mBL, tBL and HBL. All computations are carried out in R version 4.1.0 on an Intel (Core i9-10910) 3.6 GHz machine with 32 GB DDR4 memory. Let $\bm{\beta}=(3,0.5,1,1.5,1,0,\dots,0)^{\top}$ and the data generating process is based on the Model 1 in Subsection \ref{subsec:4.3}. Figure \ref{fig:CPU} shows the result of CPU times (in second) for $n = 200$ and varying dimension $p \in \{5,10,20,50,100\}$, averaged over $10$ runs. For each step, we generated 10000 posterior samples after discarding the first 5000 posterior samples as burn-in. Since we use the scale mixtures of normal representations for sampling models except for BL, computational costs are higher than that of BL as the dimension gets higher. Interestingly, the HBL is shown to have almost the same computational time as mBL and tBL, even though it contains steps to approximate the full conditional distribution.
\end{rem}

\subsection{Sensitivity analysis of hyper-parameters}
\label{subsec:4.4}

\begin{figure}[htbp]
\centering
\includegraphics[width=13cm]{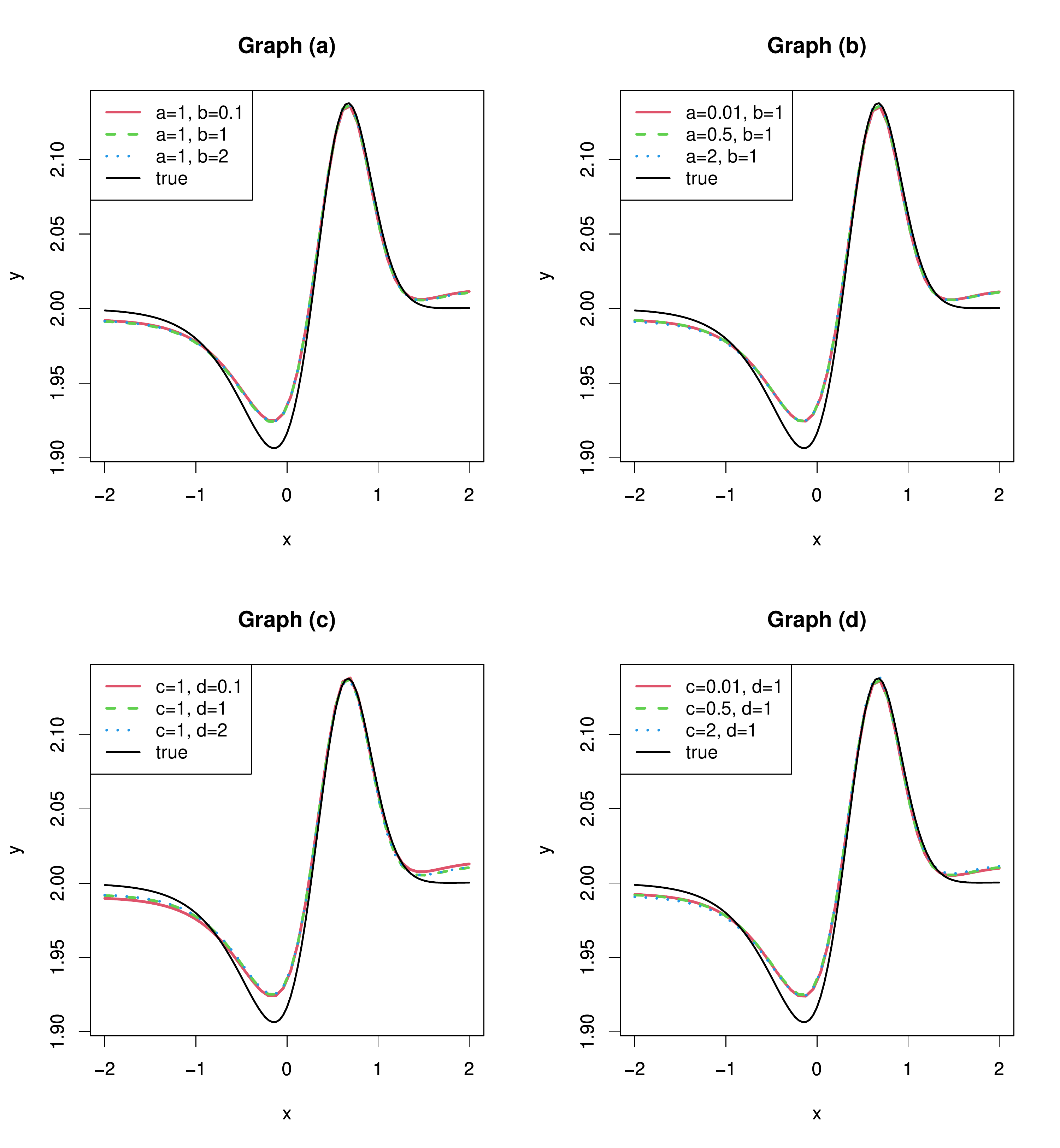}
\caption{Sensitivity analysis of hyper-parameters}
\label{fig:sensitivity}
\end{figure}

We test the sensitivity of hyper-parameters of the gamma prior for $\lambda^2$ and $\eta$ in the proposed method (model \eqref{proposal} and Algorithm \ref{algo:Gibbs2}). Following \cite{CS21}, we equally divide $x \in [-2,2]$ into $50$ pieces and the data are generated from
\begin{align*}
y_i=\bm{x}_i^{\top} \bm{\beta}+\sigma \bm{\varepsilon}, \quad \epsilon_1,\dots,\epsilon_n \overset{\mathrm{iid}}{\sim} \mathrm{Hyp}(0 , 1 , 1 , 0),\quad \sigma=0.03
\end{align*}
for $i=1,\dots, 50$ with $\bm{x}_i=((1+e^{-4(x_i-0.3)})^{-1}, (1+e^{3(x_i-0.2)})^{-1}, (1+e^{-4(x_i-0.7)})^{-1}, (1+e^{5(x_i-0.8)})^{-1})^{\top}$ and $\bm{\beta}=(1,1,1,1)^{\top}$ (see also \cite{JL07}). We consider the model \eqref{proposal} and adopt Algorithm \ref{algo:Gibbs2} to estimate $\bm{\beta}$. Note that there are four prior hyper-parameters $a,b,c,d>0$ in \eqref{proposal}, and we mainly use $a=b=c=d=1$ in this paper. We generated 3000 posterior samples after discarding the first 1000 posterior samples as burn-in, and we plot $\hat{y}_i =\bm{x}_i^{\top} \hat{\bm{\beta}}$ for $i=1,\dots,50$ in Figure \ref{fig:sensitivity}, where $\hat{\bm{\beta}}$ is the posterior mean for the Bayesian Huberized lasso using Algorithm \ref{algo:Gibbs2}. In Graphs (a) and (b), we keep $a = 1$ fixed with $b$ varied and $b = 1$ fixed with a varied $a$, respectively. For both cases, we set $c=d=1$. On the other hand, in Graph (c) and (d), we keep $c = 1$ fixed with $d$ varied and $d = 1$ fixed with a varied $c$, respectively. From the figures, we can find that the estimation results do not change very much for various selections of hyper-parameters. 

\section{Real data analysis}
\label{sec:5}

\begin{figure}[htbp]
\centering
\includegraphics[width=16cm]{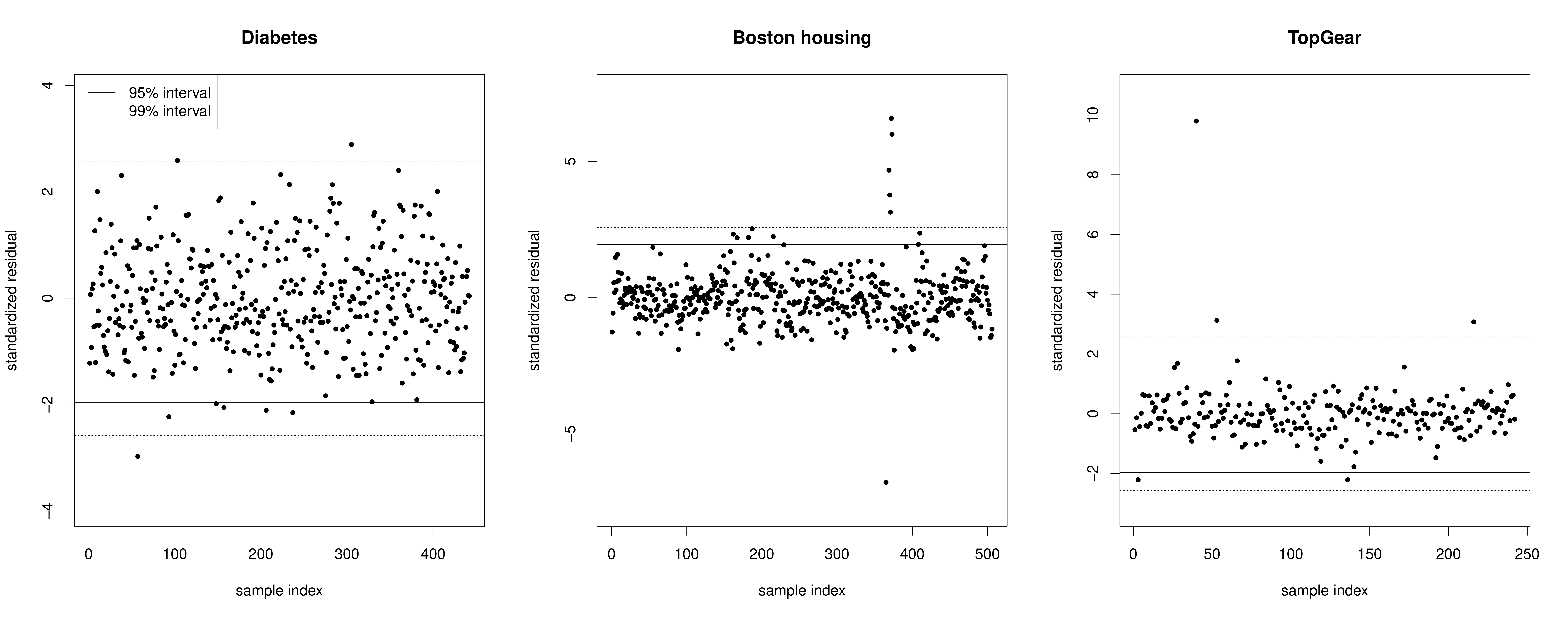}
\caption{Standardized residuals for three datasets.}
\label{fig:resid}
\end{figure}

The robustness and efficiency of the proposed Bayesian Huberized lasso (HBL) are demonstrated via the analysis of three famous datasets: Diabetes data (\cite{EHJT04}), Boston housing data (\cite{HR78}) and TopGear data (\cite{ACG16}, \cite{A21}). The standardized residual plots for three datasets are shown in Figure \ref{fig:resid}. The Diabetes data may support the standard normal assumption, while we can find that other two datasets have large outliers. In Subsection \ref{subsec:Boston} and \ref{subsec:TopGear}, we use the original data and data without outliers whose absolute values of standardized residual are  larger than 95\% interval in Figure \ref{fig:resid} to compare the effect of outliers. To help with
the interpretation of the parameters and to put the regressors on a
common scale, we assume that all of the variables are centered and scaled so
that $\bm{y}$ and the columns of $X$ all have
mean 0 and variance 1. We now compare the four methods which are the same as Subsection \ref{subsec:4.3}. In all the methods, we generated 10000 posterior samples after discarding the first 5000 posterior samples as burn-in, and we report posterior medians of regression coefficients and 95\% credible intervals. 

We also calculate the prediction error for three datasets. Since data may contain some outliers, we adopt the following four criteria as measures of predictive accuracy: mean squared prediction error (MSPE), mean absolute prediction error (MAPE), mean Huber prediction error (MHPE) for $c=1.345$ and median of squared prediction error (MedSPE) via leave-one-out cross-validation. They are defined by $\mathrm{MSPE}=n^{-1} \sum_{i=1}^n (y_i - \bm{x}_i^{\top} \hat{\bm{\beta}}^{(-i)})^2$, $\mathrm{MAPE}=n^{-1}\sum_{i=1}^n |y_i - \bm{x}_i^{\top} \hat{\bm{\beta}}^{(-i)}|$, $\mathrm{MHPE}=n^{-1}\sum_{i=1}^n \rho_c (y_i - \bm{x}_i^{\top} \hat{\bm{\beta}}^{(-i)})$ and $\mathrm{MedSPE}=\mathrm{median}_{1\le i \le n} (y_i - \bm{x}_i^{\top} \hat{\bm{\beta}}^{(-i)})^2$, where $\rho_c(\cdot)$ is the Huber loss function with $c=1.345$ and $\hat{\bm{\beta}}^{(-i)}$ is the posterior median based on dataset except for $i$-th observation. The MedMSPE is also known as the median cross validation (\cite{ZY98}). For each step, we generated 10000 posterior samples after discarding the first 5000 posterior samples as burn-in.

\subsection{Diabetes data}
\label{subsec:Diabetes}

First of all, we consider Diabetes data (\cite{EHJT04}). The data contains information of 442 individuals and 10 covariates regarding the personal information and related medical measures of the individuals, as well as a measure $y$ of
disease progression taken one year after the baseline measurements. Following \cite{HIS20},  while a regression model with ten
variables would not be overwhelmingly complex, it is suspected that
the relationship between $y$ and the $x_{j}$'s may not be linear,
and that including second-order terms like $x_{j}^{2}$ and
$x_{j}x_{k}$ in the regression model might aid in prediction. The
regressors therefore include 10 main effects $x_{1},
\dots, x_{10}, \binom{10}{2} =45$ interactions of the form $x_{j}x_{k}$ and 9 quadratic
terms $x_{j}^{2}$ (one of the regressors, $x_{2}= \mathrm{sex}$, is
binary, so $x_{2}=x_{2}^{2}$, making it unnecessary to include
$x_{2}^{2}$). This gives a total of $p=64$ regressors. 

\begin{figure}[htbp]
\centering
\includegraphics[width=15cm]{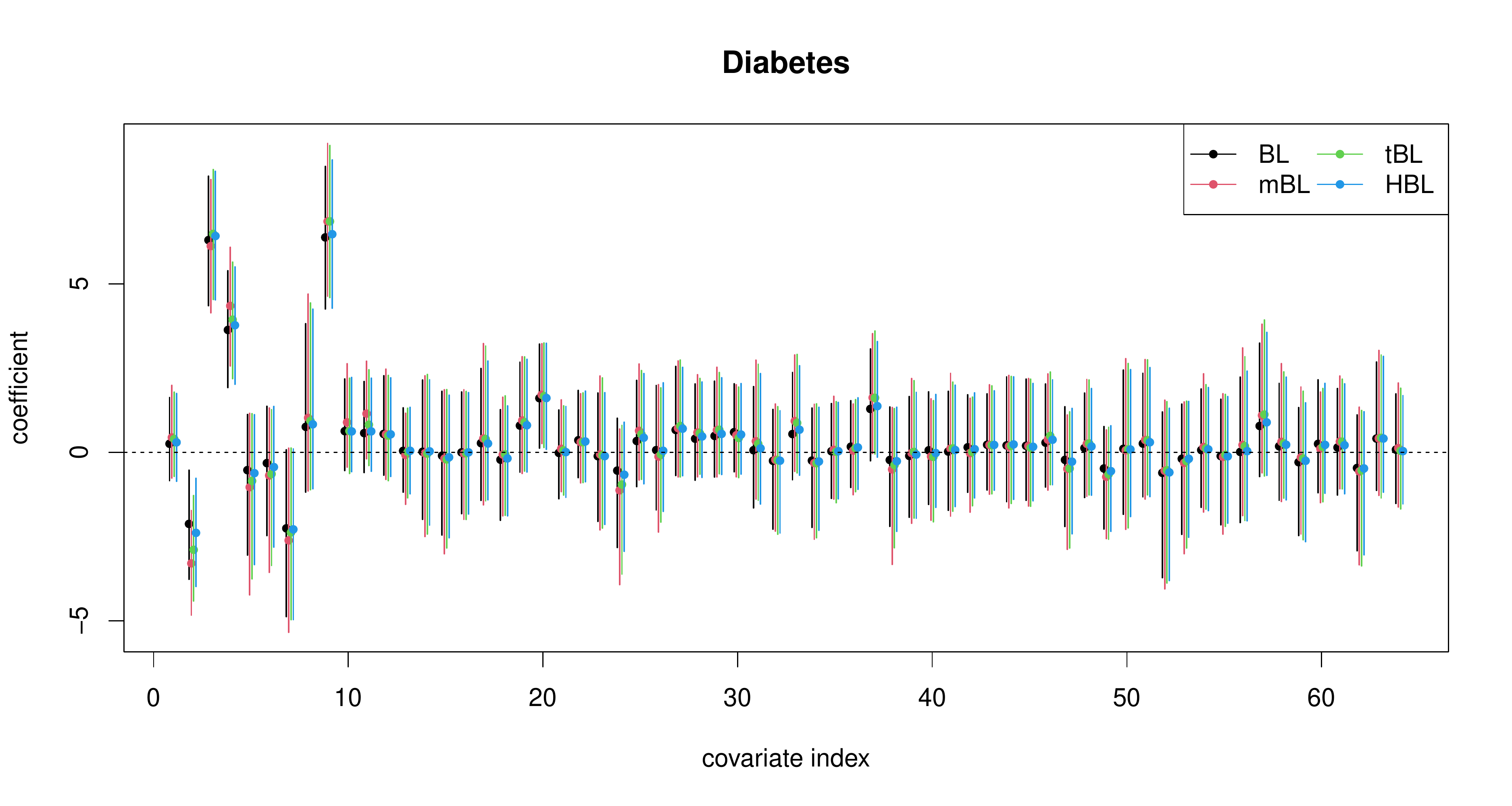}
\caption{Posterior medians and 95\% credible intervals of the regression coefficients in the original Bayesian lasso (BL), Bayesian median lasso (mBL), Bayesian lasso with $t_3$-error (tBL) and the proposed Huberized Bayesian lasso (HBL), applied to the Diabetes data.}
\label{fig:diabetes_CI}
\end{figure}

\begin{figure}[htbp]
\centering
\includegraphics[width=13cm]{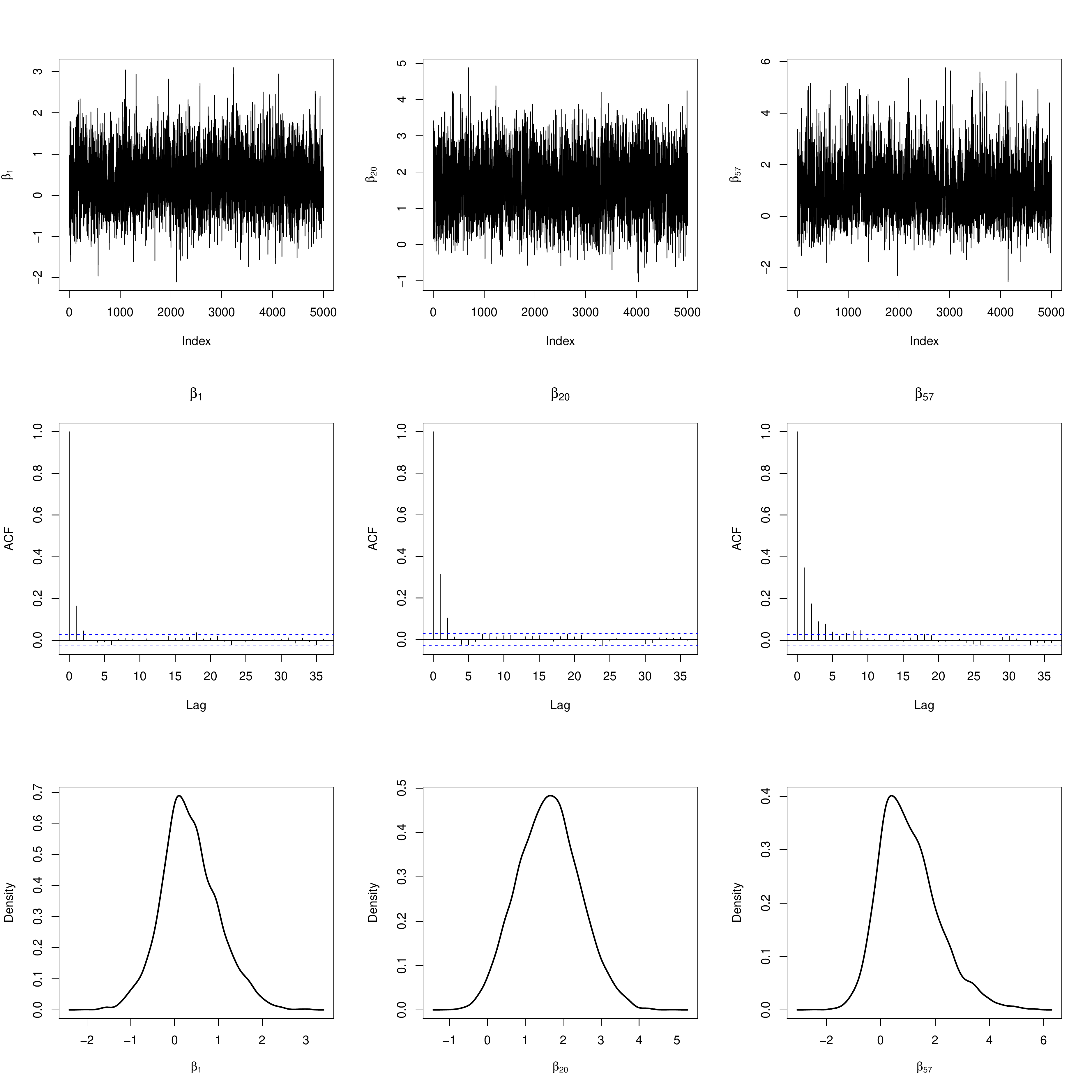}
\caption{Trajectory (top), autocorrelation (middle) and density (bottom) plots of some regression coefficients ($\beta_1$, $\beta_{20}$ and $\beta_{57}$) for HBL in Diabetes data.}
\label{fig:diabetes_MCMC}
\end{figure}

Few outliers are confirmed in the dataset as most of residuals are contained in the 99\% interval, which strongly supports the standard normal assumption in this example. Figure \ref{fig:diabetes_CI} shows the 95\% credible interval for four methods. Performances of these methods are comparable because there is no large outliers in the Diabetes dataset. In Figure \ref{fig:diabetes_MCMC}, we also report the mixing and autocorrelation plot of posterior samples based on the HBL for some regression coefficients. From trajectory of Markov chain, the mixing is reasonable and autocorrelations rapidly decay. Hence, the proposed method also seems to work well in terms of sampling from posterior distribution.   Furthermore, the average of effective sample size of posterior samples for regression coefficients was $3419.87$. The prediction performance is also shown in Table \ref{tab:diabetes}. All methods are comparable performance for four criteria.

\begin{table}[ht]
\caption{Mean squared prediction error (MSPE), mean absolute prediction error (MAPE), mean Huber prediction error (MHPE) for $c=1.345$ and median of squared prediction error (MedSPE) for Diabetes data, computed from leave-one-out cross-validation}
\begin{center}
\begin{tabular}{rrrrr}
  \hline
 & MSPE & MAPE & MHPE & MedSPE \\ 
  \hline
BL & 0.504 & 0.575 & 0.249 & 0.247 \\ 
  mBL & 0.511 & 0.576 & 0.253 & 0.253 \\ 
  tBL & 0.508 & 0.576 & 0.251 & 0.258 \\ 
  HBL & 0.504 & 0.576 & 0.250 & 0.250 \\ 
   \hline
\end{tabular}
\end{center}
\label{tab:diabetes}
\end{table}

\subsection{Boston housing data}
\label{subsec:Boston}

We compare results of the proposed method with those of the non-robust and robust Bayesian lasso through applications to Boston Housing \citep{HR78} which is a famous  dataset in investigating the normality assumption of residuals for robust estimation methods. 
The response variable in the Boston housing data is corrected median value of owner-occupied homes in USD 1000's, and there are 15 covariates including one binary covariate. 
Similar to \cite{HS20} and \cite{HIS20}, we standardized 14 continuous valued covariates, and included squared values of these covariates, which results in 29 predictors in our models. We also centered response variables.
The sample size is $506$.

\begin{figure}[htbp]
\centering
\includegraphics[width=15cm]{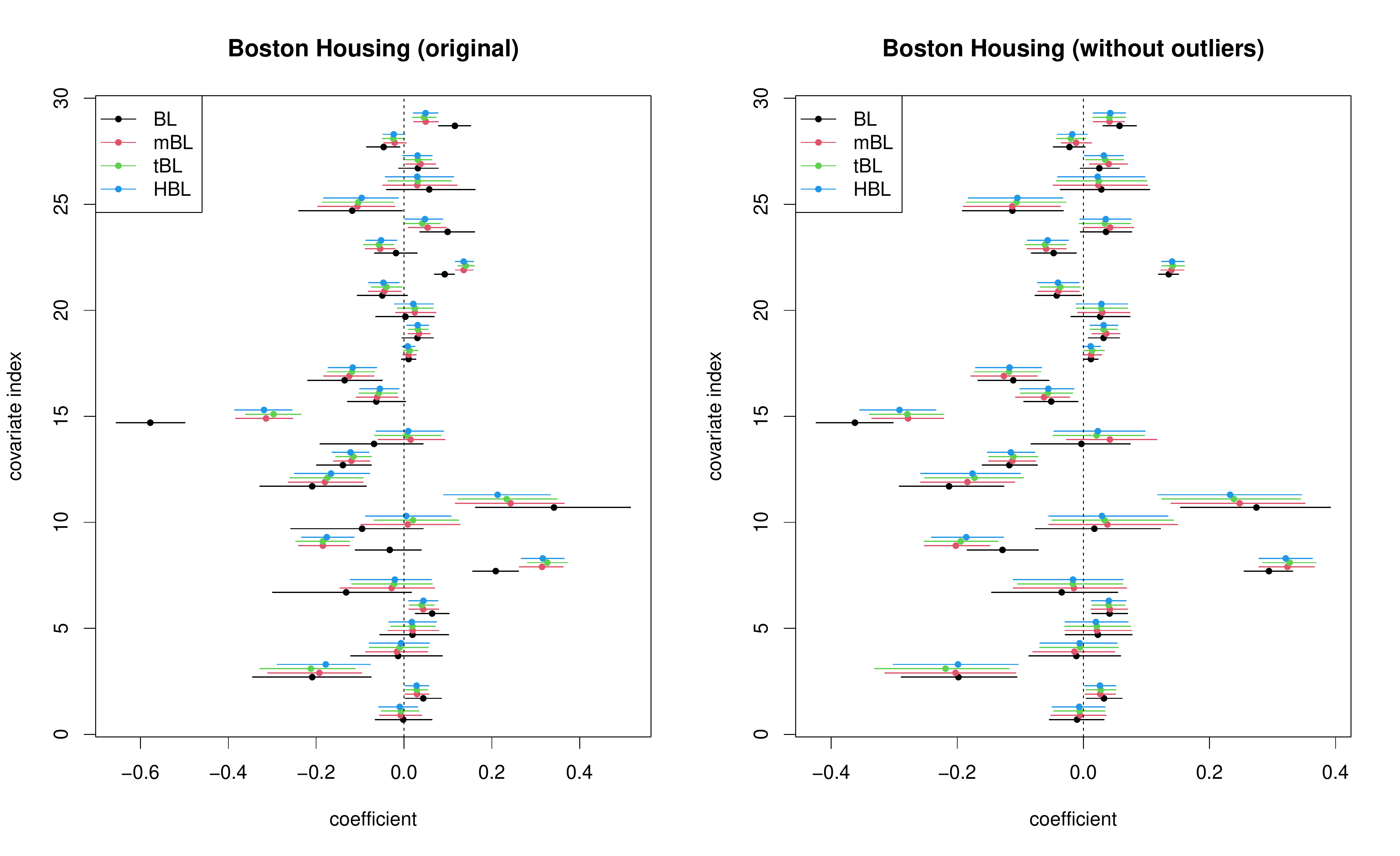}
\caption{Posterior medians and 95\% credible intervals of the regression coefficients in the original Bayesian lasso (BL), Bayesian median  lasso (mBL), Bayesian lasso with $t_3$-error (tBL) and the proposed Huberized Bayesian lasso (HBL), applied to the original Boston housing data (left) and the Boston housing data without outliers (right).}
\label{fig:boston_CI}
\end{figure}

\begin{figure}[htbp]
\centering
\includegraphics[width=13cm]{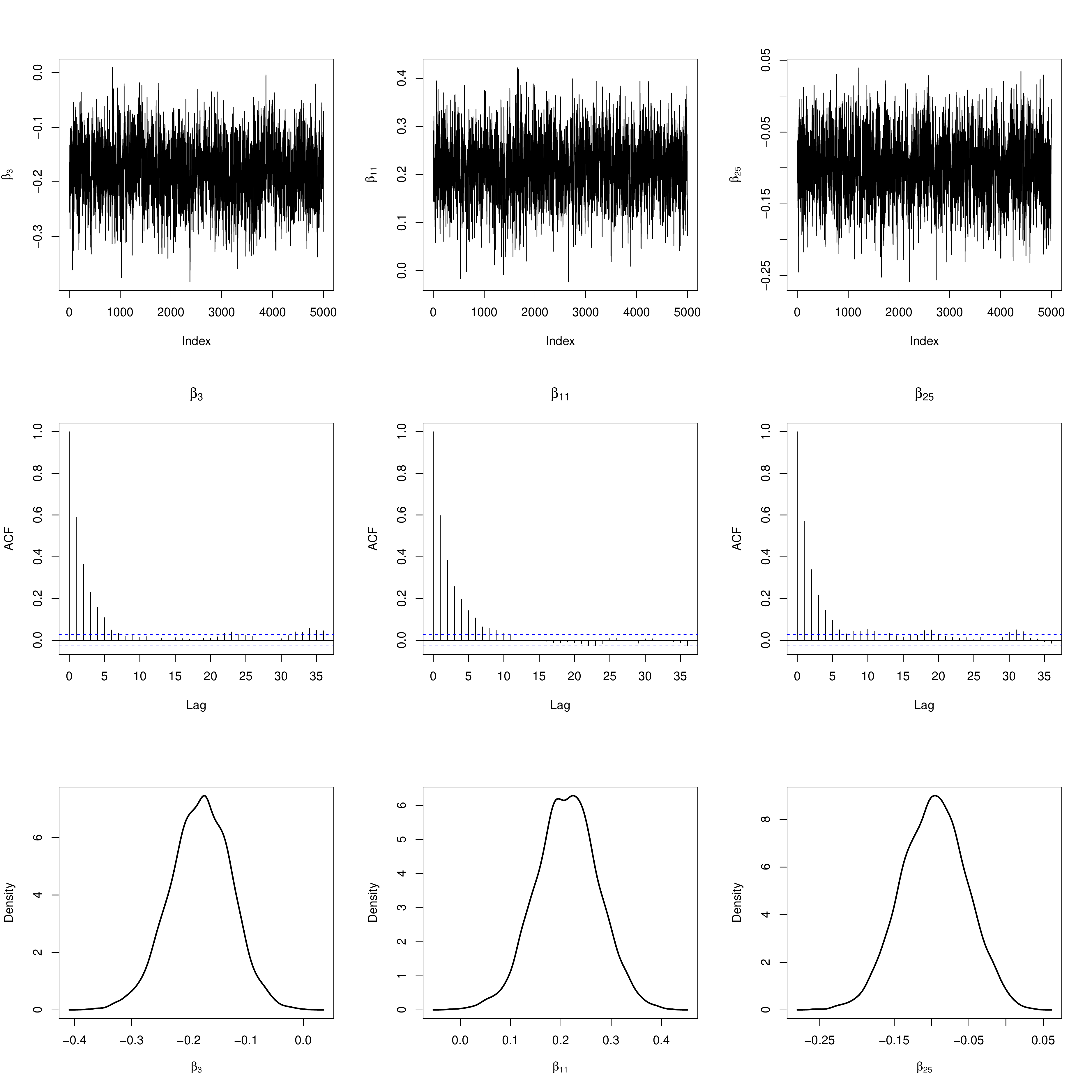}
\caption{Trajectory (top), autocorrelation (middle) and density (bottom) plots of some regression coefficients ($\beta_3$, $\beta_{11}$ and $\beta_{25}$) for HBL in Boston housing data.}
\label{fig:boston_MCMC}
\end{figure}

The posterior medians and 95\% credible intervals of the regression coefficients based on the four methods are reported in Figure \ref{fig:boston_CI}. It shows that the results of BL are quite different from other three methods, while the three robust methods, mBL, tBL and HBL produce similar results. For example, from the left panel of Figure \ref{fig:boston_CI}, the three robust methods detected 9th, 19th and 23th variables as significant ones based on their credible intervals while the credible intervals of the BL method contain 0, which shows that the robust methods may be able to detect significant variables that the non-robust method cannot. On the other hand, the right panel of Figure \ref{fig:boston_CI} shows that all methods are almost comparable in the absence of outliers. Hence robust methods including the proposed method can efficiently remove the effect of outliers. In Figure \ref{fig:boston_MCMC}, we also report the mixing and autocorrelation plot of posterior samples based on the HBL for some regression coefficients. They are also reasonable, and the average of effective sample size of posterior samples for regression coefficients was $1389.468$. For prediction errors, we can find that the HBL is the best or second place among four methods from Table \ref{tab:boston}.

\begin{table}[ht]
\caption{Mean squared prediction error (MSPE), mean absolute prediction error (MAPE), mean Huber prediction error (MHPE) for $c=1.345$ and median of squared prediction error (MedSPE) for Boston housing data, computed from leave-one-out cross-validation}
\begin{center}
\begin{tabular}{rrrrr}
  \hline
 & MSPE & MAPE & MHPE & MedSPE \\ 
  \hline
BL & 0.191 & 0.292 & 0.086 & 0.046 \\ 
  mBL & 0.210 & 0.273 & 0.089 & 0.034 \\ 
  tBL & 0.214 & 0.271 & 0.090 & 0.033 \\ 
  HBL & 0.210 & 0.272 & 0.089 & 0.031 \\ 
   \hline
\end{tabular}
\end{center}
\label{tab:boston}
\end{table}

\subsection{TopGear data}
\label{subsec:TopGear}

Finally, we use information on cars that we scraped from the website of the popular BBC television show Top Gear. The data set is included in the R package {\tt robustHD} (\cite{A21}) and contains $n = 242$ observations on $p = 29$ numerical and categorical variables. There are 4 categorical variables with two possible outcomes, and 12 categorical variables having three levels. A description of the variables is provided in Table 3 of the paper \cite{ACG16}. We use variable MPG (fuel consumption) as the response and the remaining variables as predictors. The resulting design matrix consists of 12 numerical variables, 4 individual dummy variables, and 12 groups of two dummy variables each, giving a total of 40 individual covariates. Note that we log-transform the variable Price (list price) to remove skewness. 

\begin{figure}[htbp]
\centering
\includegraphics[width=15cm]{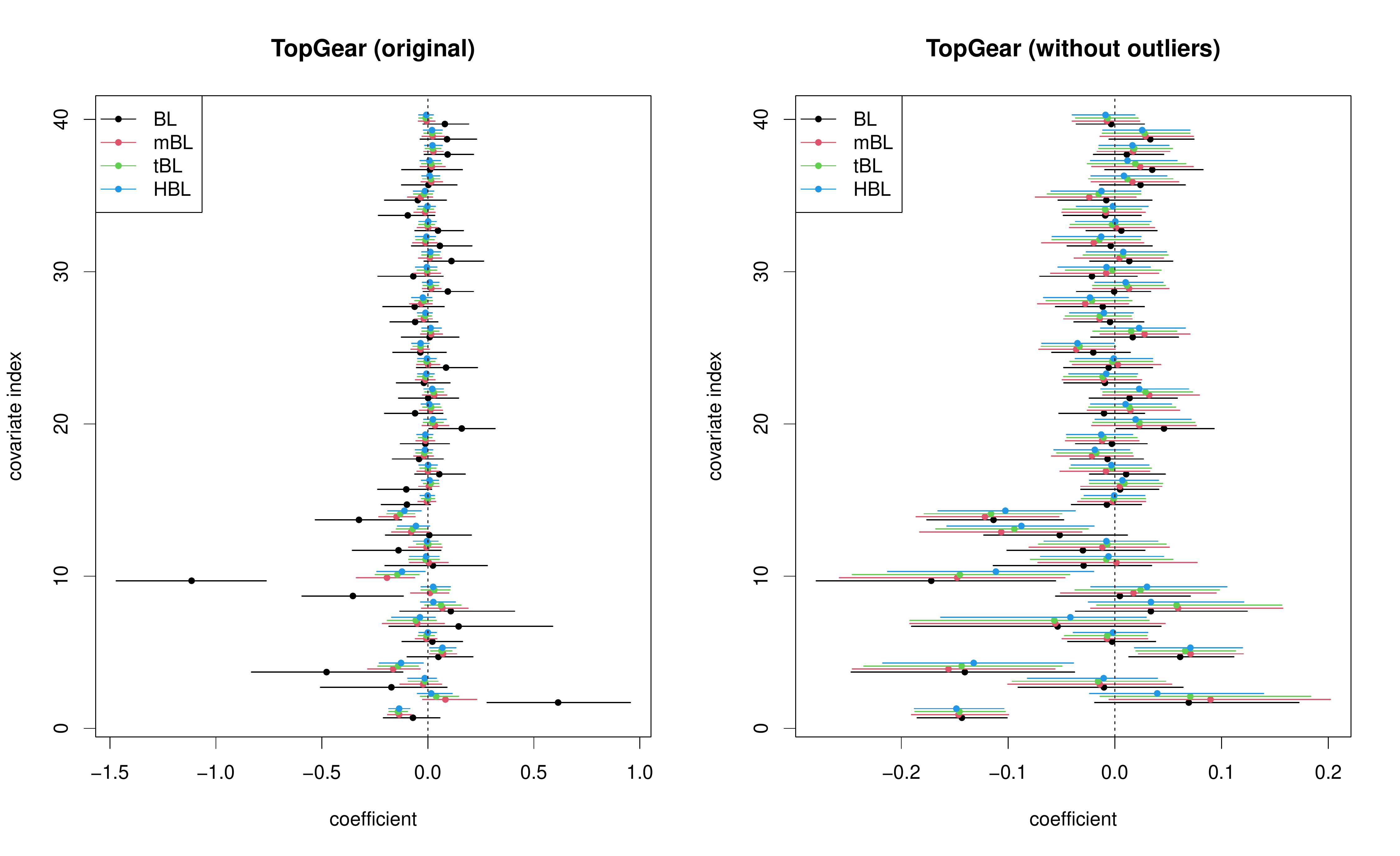}
\caption{Posterior medians and 95\% credible intervals of the regression coefficients in the original Bayesian lasso (BL), Bayesian median lasso (mBL), Bayesian lasso with $t_3$-error (tBL) and the proposed Huberized Bayesian lasso (HBL), applied to the original TopGear data (left) and the TopGear data without outliers (right).}
\label{fig:topgear_CI}
\end{figure}

\begin{figure}[htbp]
\centering
\includegraphics[width=13cm]{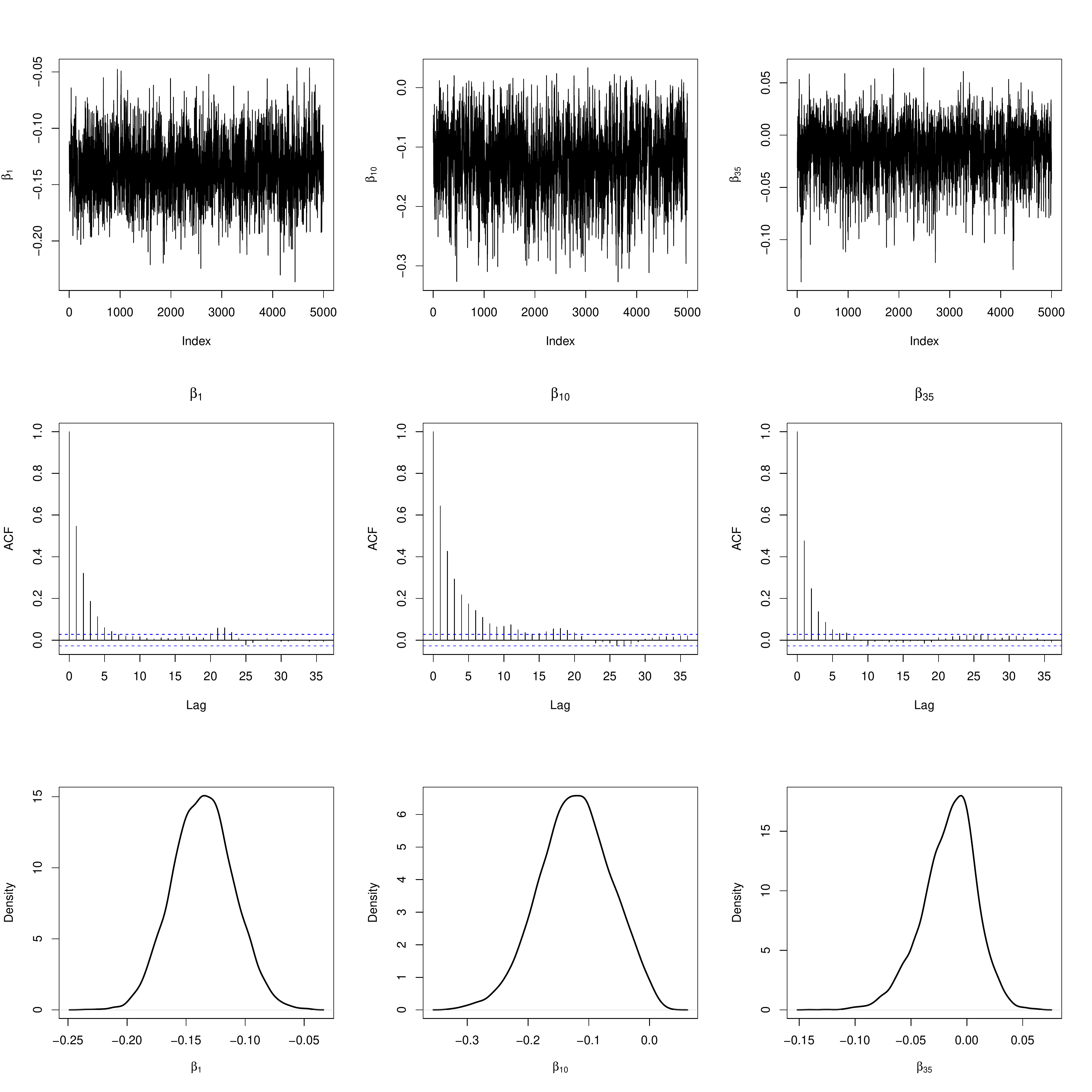}
\caption{Trajectory (top), autocorrelation (middle) and density (bottom) plots of some regression coefficients ($\beta_1$, $\beta_{10}$ and $\beta_{35}$) for HBL in TopGear data.}
\label{fig:topgear_MCMC}
\end{figure}

As mentioned in \cite{ACG16}, it reveals three clear outliers from the standardized residual plot in the right panel of Figure \ref{fig:resid}: BMW i3 (observation 40), Chevrolet Volt (observation 53) and Vauxhall Ampera (observation 216). All three are electric cars with an additional petrol-powered range extender engine. The posterior medians and 95\% credible intervals of the regression coefficients based on the four methods are reported in Figure \ref{fig:topgear_CI}. As Boston housing data, it also shows that the results of BL are quite different from other three methods. We also report the mixing and autocorrelation plot of posterior samples based on the HBL for some regression coefficients in Figure \ref{fig:topgear_MCMC}. From trajectory of Markov chain, the mixing is reasonable and autocorrelations rapidly decay. Furthermore, the average of effective sample size of posterior samples for regression coefficients was $1705.048$. From Table \ref{tab:topgear}, predictive performance of three robust methods are comparable and non-robust Bayesian lasso is worse than the others. We note that \cite{ACG16} presented the data analysis for TopGear data using robust group lasso because the TopGear data contains many categorical variables. To select interpretable variables, the proposed method should be extended to the Bayesian Huberized group lasso in the future.

\begin{table}[ht]
\caption{Mean squared prediction error (MSPE), mean absolute prediction error (MAPE), mean Huber prediction error (MHPE) for $c=1.345$ and median of squared prediction error (MedSPE) for TopGear data, computed from leave-one-out cross-validation}
\begin{center}
\begin{tabular}{rrrrr}
  \hline
 & MSPE & MAPE & MHPE & MedSPE \\ 
  \hline
BL & 0.814 & 0.430 & 0.202 & 0.082 \\ 
  mBL & 0.775 & 0.281 & 0.135 & 0.025 \\ 
  tBL & 0.787 & 0.280 & 0.136 & 0.027 \\ 
  HBL & 0.797 & 0.279 & 0.136 & 0.028 \\ 
   \hline
\end{tabular}
\end{center}
\label{tab:topgear}
\end{table}

\section{Concluding remarks}
\label{sec:6}

We proposed a posterior computation algorithm which involves the estimation of a tuning parameter for robustness in the Bayesian Huberized lasso model. The proposed Gibbs sampler is based on the approximation of the intractable full conditional distribution by using the similar idea from \cite{M19}. Through some simulation studies, it was shown that the proposed method does not have the strong robustness like the Bayesian lasso with t-error (tBL), but shows stable performance with or without outliers.

We close this paper by considering future directions. First, it is interesting to study some theoretical properties of the proposed algorithm. The proposed Gibbs sampler is not exact MCMC but approximate one. Although the proposed method numerically seems to derive reasonable estimates for regression coefficient vector, it is not shown whether the corresponding posterior theoretically converges to the target posterior distribution or not. On the other hand, it is interesting to show the geometric ergodicity of the Gibbs sampler for the Bayesian Huberized lasso for fixed $\eta$. For example, \cite{KH13} proved the geometric ergodicity of the Gibbs sampler for original Bayesian lasso. Second, although the estimation of mean function or parameter is important, the robust estimation of the covariance matrix is also important. The Bayesian Huberized lasso should be extend to the Bayesian graphical lasso (e.g. \cite{W12}) in the future. Finally, it is important to consider other selection methods of $\eta$. Although this paper proposed the estimation of $\eta$ in fully Bayesian framework, the optimal tuning parameter in terms of prediction is also important. In particular, the predictive covariance information criterion (PCIC) recently proposed by \cite{IY21} is based on the quasi-Bayes posterior and the advantage of the PCIC is to calculate by using only one-shot MCMC output. However, in the presence of outliers, the selection of weights in the criterion will be quite challenging.

\section*{Acknowlegements}

This work is partially supported by Japan Society for Promotion of Science (KAKENHI) grant numbers 21K13835.

\appendix

\section{Proofs}

In this section, we give the proofs of Propositions \ref{propriety}, \ref{unimodality} and \ref{fixed_point}. 

\subsection*{Proof of Proposition \ref{propriety}}

The overall posterior distribution is given by
\begin{align*}
\pi(\bm{\beta},\bm{\sigma}^{2},\bm{\tau}^{2},\rho^{2}\mid\bm{y})=\frac{f(\bm{y}\mid X,\bm{\beta},\bm{\sigma}^{2})\pi(\bm{\beta}\mid\bm{\tau}^{2},\rho^{2})\pi({\tau}^{2})\pi(\bm{\sigma}^{2}\mid\rho^{2})\pi(\rho^{2})}{\int f(\bm{y}\mid X,\bm{\beta},\bm{\sigma}^{2})\pi(\bm{\beta}\mid\bm{\tau}^{2},\rho^{2})\pi({\tau}^{2})\pi(\bm{\sigma}^{2}\mid\rho^{2})\pi(\rho^{2})d\bm{\beta}d\bm{\sigma}^{2}d\bm{\tau}^{2}d\rho^{2}}.
\end{align*}
We show that the normalizing constant of the posterior distribution is finite, that is, 
\begin{align*}
\int f(\bm{y}\mid X,\bm{\beta},\bm{\sigma}^{2})\pi(\bm{\beta}\mid\bm{\tau}^{2},\rho^{2})\pi({\tau}^{2})\pi(\bm{\sigma}^{2}\mid\rho^{2})\pi(\rho^{2})d\bm{\beta}d\bm{\sigma}^{2}d\bm{\tau}^{2}d\rho^{2}<\infty.
\end{align*}
First, we consider the integral with respect to $\bm{\beta}$. We have
\begin{align*}
\int f(\bm{y}\mid X,\bm{\beta},\bm{\sigma}^{2})\pi(\bm{\beta}\mid\bm{\tau}^{2},\rho^{2})d\bm\beta&=\int(2\pi)^{-n/2}\left(\prod_{i=1}^{n}\sigma_{i}^{2}\right)^{-1/2}\exp\left\{-\dfrac{1}{2}({\bm{y}}-X\bm\beta)^{\top}D_{\bm\sigma}^{-1}({\bm{y}}-X\bm\beta)\right\}\\
&\times (2\pi)^{-p/2}(\rho^{2})^{-p/2}|D_{\bm\tau}|^{-1/2}\exp\left(-\dfrac{1}{2\rho^{2}}\bm{\beta}^{\top}D_{\bm\tau}^{-1}\bm\beta\right)d\bm\beta\\
&=(2\pi)^{-(n+p)/2}(\rho^{2})^{-p/2}\left(\prod_{i=1}^{n}\sigma_{i}^{2}\right)^{-1/2}\left(\prod_{j=1}^{p}\tau_{j}^{2}\right)^{-1/2}\\
&\times \int\exp\left\{-\dfrac{1}{2}({\bm{y}}-X\bm\beta)^{\top}D_{\bm\sigma}^{-1}({\bm{y}}-X\bm\beta)-\dfrac{1}{2\rho^{2}}\bm{\beta}^{\top}D_{\bm\tau}^{-1}\bm\beta\right\}d\bm\beta.
\end{align*}
In particular, we have
\begin{align*}
&\int\exp\left\{-\dfrac{1}{2}({\bm{y}}-X\bm\beta)^{\top}D_{\bm\sigma}^{-1}({\bm{y}}-X\bm\beta)-\dfrac{1}{2\rho^{2}}\bm{\beta}^{\top}D_{\bm\tau}^{-1}\bm\beta\right\}d\bm\beta\\
&=\exp\left(-\dfrac{1}{2}{\bm{y}}^{\top}D_{\bm\sigma}^{-1}{\bm{y}}\right)\int\exp\left\{-\dfrac{1}{2}\bm{\beta}^{\top}\left(X^{\top}D_{\bm\sigma}^{-1}X+\dfrac{1}{\rho^{2}}D_{\bm\tau}^{-1}\right)\bm{\beta}+\bm{\beta}^{\top}X^{\top}D_{\bm\sigma}^{-1}{\bm{y}}\right\}d\bm\beta\\
&=\exp\left(-\dfrac{1}{2}{\bm{y}}^{\top}D_{\bm\sigma}^{-1}{\bm{y}}\right)(2\pi)^{p/2}\left|\left(X^{\top}D_{\bm\sigma}^{-1}X+\dfrac{1}{\rho^{2}}D_{\bm\tau}^{-1}\right)^{-1}\right|^{1/2}\\
&=\exp\left(-\dfrac{1}{2}{\bm{y}}^{\top}D_{\bm\sigma}^{-1}{\bm{y}}\right)(2\pi)^{p/2}\left|\dfrac{1}{\rho^{2}}D_{\bm\tau}^{-1}\right|^{-1/2}|D_{\bm{\sigma}}|^{-1/2}|D_{\bm{\sigma}}+\rho^{2}XD_{\bm\tau}X^{\top}|^{-1/2}\\
&=\exp\left(-\dfrac{1}{2}{\bm{y}}^{\top}D_{\bm\sigma}^{-1}{\bm{y}}\right)(2\pi)^{p/2}(\rho^{2})^{p/2}\left(\prod_{j=1}^{p}\tau_{j}^{2}\right)^{1/2}\left(\prod_{i=1}^{n}\sigma_{i}^{2}\right)^{1/2}|D_{\bm{\sigma}}+\rho^{2}XD_{\bm\tau}X^{\top}|^{-1/2}.
\end{align*}
Hence, we have
\begin{align*}
\int f(\bm{y}\mid X,\bm{\beta},\bm{\sigma}^{2})\pi(\bm{\beta}\mid\bm{\tau}^{2},\rho^{2})d\bm\beta=(2\pi)^{-n/2}|D_{\bm{\sigma}}+\rho^{2}XD_{\bm\tau}X^{\top}|^{-1/2}\exp\left(-\dfrac{1}{2}{\bm{y}}^{\top}D_{\bm\sigma}^{-1}{\bm{y}}\right).
\end{align*}
Next, we have
\begin{align*}
&\int f(\bm{y}\mid X,\bm{\beta},\bm{\sigma}^{2})\pi(\bm{\beta}\mid\bm{\tau}^{2},\rho^{2})\pi({\tau}^{2})\pi(\bm{\sigma}^{2}\mid\rho^{2})\pi(\rho^{2})d\bm{\beta}d\bm{\sigma}^{2}d\bm{\tau}^{2}d\rho^{2}\\
=&\iiint\left\{(2\pi)^{-n/2}|D_{\bm{\sigma}}+\rho^{2}XD_{\bm\tau}X^{\top}|^{-1/2}\exp\left(-\dfrac{1}{2}{\bm{y}}^{\top}D_{\bm\sigma}^{-1}{\bm{y}}\right)\right\}\\
&\times \pi(\bm{\tau}^{2})\pi(\bm{\sigma}^{2}\mid\rho^{2})\pi(\rho^{2})d\bm{\sigma}^{2}d\bm{\tau}^{2}d\rho^{2}\\
\leq& (2\pi)^{-n/2}\iiint|D_{\bm{\sigma}}|^{-1/2}\exp\left(-\dfrac{1}{2}{\bm{y}}^{\top}D_{\bm\sigma}^{-1}{\bm{y}}\right) \prod_{i=1}^{n}\dfrac{1}{2\rho^2 K_1(\eta)} \exp\left\{-\dfrac{\eta}{2}\left(\dfrac{\sigma_i^2}{\rho^2}+\dfrac{\rho^2}{\sigma_i^2}\right)\right\}\\\
&\times\pi(\bm{\tau}^{2})\pi(\rho^{2})d\bm{\sigma}^{2}d\bm{\tau}^{2}d\rho^{2}
\end{align*}
by using the fact that $|A+B|\geq|A|$ implies $|A+B|^{-1/2}\leq|A|^{-1/2}$ for a positive definite matrix  $A$ and a semi-positive definite matrix $B$. 

Next, we consider the integral with respect to $\bm{\sigma}^2$. First, we have
\begin{align*}
&\int|D_{\bm{\sigma}}|^{-1/2}\exp\left(-\dfrac{1}{2}\bm{y}^{\top}D_{\bm\sigma}^{-1}\bm{y}\right) \prod_{i=1}^{n}\dfrac{1}{2\rho^2 K_1(\eta)} \exp\left\{-\dfrac{\eta}{2}\left(\dfrac{\sigma_i^2}{\rho^2}+\dfrac{\rho^2}{\sigma_i^2}\right)\right\}d\bm\sigma^{2}\\
&=\int\prod_{i=1}^{n}(\sigma^{2}_{i})^{-1/2}\exp\left(-\dfrac{1}{2}\dfrac{y_{i}^{2}}{\sigma^{2}_{i}}\right) \dfrac{1}{2\rho^2 K_1(\eta)} \exp\left\{-\dfrac{\eta}{2}\left(\dfrac{\sigma_i^2}{\rho^2}+\dfrac{\rho^2}{\sigma_i^2}\right)\right\}d\bm\sigma^{2}\\
&=\left(\dfrac{1}{2\rho^2 K_1(\eta)}\right)^{n}\int\prod_{i=1}^{n}(\sigma^{2}_{i})^{1/2-1}\exp\left\{-\dfrac{1}{2}\left(\dfrac{\eta}{\rho^2}\sigma_i^2+\dfrac{y_{i}^{2}+\eta\rho^2}{\sigma_i^2}\right)\right\}d\bm\sigma^{2}.
\end{align*}
Letting $a=\eta/\rho^{2}$ and $b=y_{i}^{2}+\eta\rho^{2}$, we obtain
\begin{align*}
&\left(\dfrac{1}{2\rho^2 K_1(\eta)}\right)^{n}\int\prod_{i=1}^{n}(\sigma^{2}_{i})^{1/2-1}\exp\left\{-\dfrac{1}{2}\left(a\sigma_i^2+\dfrac{b}{\sigma_i^2}\right)\right\}d\bm\sigma^{2}\\
&=\prod_{i=1}^{n}\dfrac{(\pi)^{1/2}}{\sqrt{2\eta\rho^2} K_1(\eta)}\exp\left\{-\sqrt{\eta\left(\eta+\dfrac{y_{i}^2}{\rho^{2}}\right)}\right\}.
\end{align*}
So, we have
\begin{align*}
&\int f(\bm{y}\mid X,\bm{\beta},\bm{\sigma}^{2})\pi(\bm{\beta}\mid\bm{\tau}^{2},\rho^{2})\pi({\tau}^{2})\pi(\bm{\sigma}^{2}\mid\rho^{2})\pi(\rho^{2})d\bm{\beta}d\bm{\sigma}^{2}d\bm{\tau}^{2}d\rho^{2}\\
&\leq\iint(2\pi)^{-n/2}\prod_{i=1}^{n}\dfrac{(\pi)^{1/2}}{\sqrt{2\eta\rho^2} K_1(\eta)}\exp\left\{-\sqrt{\eta\left(\eta+\dfrac{y_{i}^2}{\rho^{2}}\right)}\right\}\prod_{j=1}^{p}\dfrac{\lambda^2}{2}e^{-\lambda^{2}\tau_{j}^2/2}\dfrac{1}{\rho^2}d\bm\tau^2 d\rho^2
\end{align*}
and 
\begin{align}\label{propriety_proof}
&\iint(2\pi)^{-n/2}\prod_{i=1}^{n}\dfrac{(\pi)^{1/2}}{\sqrt{2\eta\rho^2} K_1(\eta)}\exp\left\{-\sqrt{\eta\left(\eta+\dfrac{{y}_{i}^2}{\rho^{2}}\right)}\right\}\prod_{j=1}^{p}\dfrac{\lambda^2}{2}e^{-\lambda^{2}\tau_{j}^2/2}\dfrac{1}{\rho^2}d\bm\tau^2 d\rho^2 \notag \\
&=\int(2\pi)^{-n/2}\prod_{i=1}^{n}\dfrac{(\pi)^{1/2}}{\sqrt{2\eta\rho^2} K_1(\eta)}\exp\left\{-\sqrt{\eta\left(\eta+\dfrac{{y}_{i}^2}{\rho^{2}}\right)}\right\}\dfrac{1}{\rho^{2}}d\rho^2 \notag \\
&=\left(\dfrac{1}{\sqrt{2\eta} K_1(\eta)}\right)^{n}\int(\rho^2)^{-n/2-1}\exp\left\{-\sum_{i=1}^n\sqrt{\eta\left(\eta+\dfrac{{y}_{i}^2}{\rho^{2}}\right)}\right\}d\rho^2.
\end{align}
In \eqref{propriety_proof}, we note that the inequality 
\[\sqrt{\eta\left(\eta+\dfrac{y_{i}^2}{\rho^{2}}\right)}=\sqrt{\eta^2 +\eta \frac{y_i^2}{\rho^2}} \ge \sqrt{\eta \frac{y_i^2}{\rho^2}}\]
holds for any $\eta>0$. Hence, we have
\begin{align*}
&\int_0^{\infty}(\rho^2)^{-(n/2)-1} \exp\left\{-\sum_{i=1}^n\sqrt{\eta\left(\eta+\dfrac{y_{i}^2}{\rho^{2}}\right)}\right\}d\rho^2\\
& \le \int_0^{\infty}(\rho^2)^{-(n/2)-1} \exp\left\{-\sum_{i=1}^n \sqrt{\eta \frac{y_i^2}{\rho^2}}\right\}d\rho^2\\
&= \int_0^{\infty}(\rho^2)^{-(n/2)-1} \exp\left\{-\left(\eta\sum_{i=1}^n \sqrt{y_i^2} \right) \frac{1}{\sqrt{\rho^2}}\right\}d\rho^2\\
&= \int_0^{\infty}(\sqrt{\rho^2})^{-n-2} \exp\left\{-\left(\eta\sum_{i=1}^n \sqrt{y_i^2} \right) \frac{1}{\sqrt{\rho^2}}\right\}d\rho^2.
\end{align*}
By using the transformation $\sqrt{\rho^2} = t$, we have
\begin{align*}
\int_0^{\infty}(\sqrt{\rho^2})^{-n-2} \exp\left\{-\left(\eta\sum_{i=1}^n \sqrt{y_i^2} \right) \frac{1}{\sqrt{\rho^2}}\right\}d\rho^2=2\int_0^{\infty}t^{-n-1} \exp\left\{-\left(\eta\sum_{i=1}^n \sqrt{y_i^2} \right) \frac{1}{t}\right\}dt.
\end{align*}
Since the integrand is the kernel of $\mathrm{IG}(n, \eta\sum_{i=1}^n \sqrt{y_i^2})$, this integral is finite for any $n$.

Hence, the posterior distribution under the improper prior $\pi(\rho^2) \propto 1/\rho^2$ is proper for any $n$.

\qed

\subsection*{Proof of Proposition \ref{unimodality}}

The joint posterior density of $(\bm\beta,\rho^{2})$ is expressed by

\begin{align*}
\pi(\bm\beta,\rho^{2}\mid\bm{y})
&=\int f(\bm{y} \mid \mu,X,\bm{\beta},\bm{\sigma}^2 )\pi(\bm{\beta} \mid \rho^2)\pi(\sigma_i^2 \mid \rho^2)\pi(\rho^2)d\bm{\sigma}^{2}\\
&=\pi(\bm{\beta} \mid \rho^2)\pi(\rho^2)\times\prod_{i=1}^{n} \frac{1}{2K_1(\eta)\sqrt{\eta \rho^2}} \exp\left(-\sqrt{\eta \left\{\eta+(y_i-\bm{x}_i^{\top}\bm{\beta})^2/\rho^2\right\}}\right)\\
&\propto(\rho^{2})^{-(n+p)/2}\exp\left\{ -\frac{\eta}{\sqrt{\rho^{2}}}\sum_{j=1}^{p}|\beta_{j}|-\sum_{i=1}^{n}\sqrt{\eta \left\{\eta+(y_i-\bm{x}_i^{\top}\bm{\beta})^2/\rho^2\right\}}\right\}.
\end{align*}
Then the log-posterior density is given by
\begin{align}\label{unimodal_logpos}
\begin{split}
\log[\pi(\bm\beta,\rho^{2}\mid \bm{y})]=& -\frac{n+p}{2}\log(\rho^{2})-\frac{\eta}{\sqrt{\rho^{2}}}||\bm\beta||_{1}-\sum_{i=1}^{n-1}\sqrt{\eta \left\{\eta+\frac{(y_i-\bm{x}_i^{\top}\bm{\beta})^2}{\rho^2}\right\}}\\
&+(\text{terms not depend on $\bm\beta$ and $\rho^{2}$}).
\end{split}
\end{align}
We consider the coordinate transformation $\bm{\phi}\leftrightarrow\bm{\beta}/\sqrt{\rho^2},\quad \xi\leftrightarrow1/\sqrt{\rho^2}$.  
In the transformed coordinates, \eqref{unimodal_logpos} after dropping terms not depend on $\bm\beta$ and $\rho^{2}$ is 
\begin{align}\label{unimodal_logpos2}
(n+p)\log \xi-\eta||\bm\phi||_{1}-\sum_{i=1}^{n}\sqrt{\eta\{\eta+(\xi y_i-\bm{x}_i^{\top}\bm{\phi})^2}.
\end{align}
Since the three terms in \eqref{unimodal_logpos2} are concave, hence the joint posterior $\pi(\bm\beta,\rho^{2}\mid \bm{y})$ is unimodal. This completes the proof.

\qed

\subsection*{Proof of Proposition \ref{fixed_point}}

Althogh the first half of the proof is the almost same as that of \cite{M19}, we give the proof for the sake of completeness. We consider the following model
\begin{align*}
X_{1},\dots,X_{n}\mid\eta,\rho^2 &\overset{\mathrm{iid}}{\sim}\mathrm{GIG}({\nu},\eta,\rho^2),\quad(\nu\in\mathbb{R} , \eta>0,\ \rho^2>0)\\
\eta\mid a,b&{\sim}\mathrm{Ga}(a,b),
\end{align*} 
where $\nu\in \mathbb{R}$ and $\rho^2>0$ are fixed constants. Note that Proposition \ref{fixed_point} is the special case of $\nu=1$. Let $f$ be the true full conditional density and $g$ be the gamma $\mathrm{Ga}(A,B)$ density. 
First of all, we show that the equation $(\partial/\partial \eta)\log f(\eta)+(1/\eta)=0$ characterizes a fixed point. Let 
\begin{align*}
d_{1}(\eta)&=\frac{\partial}{\partial \eta}\log f(\eta)=-n\frac{\partial}{\partial \eta}\log K_{\nu}(\eta)+\frac{a-1}{\eta}-P-b,\\
d_{2}(\eta)&=\frac{\partial^2}{\partial \eta^2}\log f(\eta)=-n\frac{\partial^2}{\partial \eta^2}\log K_{\nu}(\eta) -\frac{a-1}{\eta^2}.
\end{align*}
For a current value $\eta$, we define $A$ and $B$ such that $d_{1}(\eta)=(\partial/\partial \eta)\log g(\eta)$ and $d_{2}(\eta)=(\partial^2/\partial \eta^2)\log g(\eta)$, and we set $\eta=A/B$ and update $\eta$. We note that it holds that
\begin{align*}
&d_{1}(\eta)=\frac{\partial}{\partial \eta}\log g(\eta)=\frac{A-1}{\eta}-B ,\quad 
d_{2}(\eta)=\frac{\partial^2}{\partial \eta^2}\log g(\eta)=-\frac{A-1}{\eta^2}.
\end{align*}
By solving the equations, we have $A=-\eta^2 d_{2}(\eta)+1$ and $B=-\eta d_{2}(\eta)-d_{1}(\eta)$. Hence, we have the following identity:
\begin{align}\label{fixed_point_proof}
\eta=\frac{A}{B}=\frac{\eta^2 d_{2}(\eta)-1}{\eta d_{2}(\eta)+d_{1}(\eta)}.
\end{align}
If it holds that $d_{1}(\eta)+(1/\eta)=0$, the right-hand side of the equation \eqref{fixed_point_proof} is 
\begin{align*}
\frac{\eta^2 d_{2}(\eta)-1}{\eta d_{2}(\eta)+d_{1}(\eta)}=\frac{\eta^2 d_{2}(\eta)-1}{\eta d_{2}(\eta)-1/\eta}=\eta.
\end{align*}
Since we have $A(\eta)/B(A(\eta),\eta)=\eta$, $\eta$ is a fixed point of Algorithm \eqref{AB}. If $\eta$ is a fixed point, we note that it must be $\eta d_{2}(\eta)+d_{1}(\eta)\neq0$. 
On the other hand, if it holds that $d_{1}(\eta)+1/\eta=0$, then we have $\eta d_{2}(\eta)+d_{1}(\eta)\neq0$, and otherwise since we have
\begin{align*}
\frac{1}{\eta^2}=d_{2}(\eta)=-n\frac{\partial^2}{\partial \eta^2}\log K_{\nu}(\eta)-\frac{a-1}{\eta^2},
\end{align*}
or equivalently,
\begin{align*}
n\frac{\partial^2}{\partial \eta^2}\log K_{\nu}(\eta)=-\frac{a}{\eta^2}<0.
\end{align*}
It is contradiction to $n(\partial^2/\partial \eta^2)\log K_{\nu}(\eta)>0$ for any $\eta>0$. In other words, it holds that $\eta d_{2}(\eta)+d_{1}(\eta)\neq0$, and if $\eta$ is a fixed point, $d_{1}(\eta)+1/\eta=0$ for any $\eta>0$. 

Next, we show that there exists $\eta>0$ such that $(\partial/\partial \eta)\log f(\eta)+(1/\eta)=0$. We note that it can be easily shown that $(\partial/\partial \eta)\log K_{\nu}(\eta)\to -\infty$ as $\eta\to0$, and 
\begin{align}\label{bessel_lim}
\frac{\partial}{\partial \eta}\log K_{\nu}(\eta)\to -1
\end{align}
as $\eta\to\infty$. \eqref{bessel_lim} is proved by using the fact $K_{\nu}(\eta){\sim}\sqrt{\pi/(2\eta)}e^{-\eta}$ as $\eta\to\infty$ (e.g. \cite{AS65}) and the formula $(d/d\eta)\log K_{\nu}(\eta)=-\{K_{\nu-1}(\eta)+K_{\nu+1}(\eta)\}/\{2K_{\nu}(\eta)\}$ for any $\nu \in \mathbb{R}$.  Finally, we will show that 
\begin{align*}
&d_{1}(\eta)+\frac{1}{\eta} =-n\frac{\partial}{\partial \eta}\log K_{\nu}(\eta)+\frac{a}{\eta}-P-b <0
\end{align*}
for large $\eta>0$, where $x_{i}>0$ for $i=1,\dots,n$, $\rho^2>0$ and $P=(1/2)\sum_{i=1}^n \{(x_{i}/\rho^2)+(\rho^2/x_{i})\} >0$. Since
\[-n\frac{\partial}{\partial \eta}\log K_{\nu}(\eta)+\frac{a}{\eta}-P-b\to n-P-b\]
as $\eta \to \infty$, it is enough to show that $n-P-b<0$ implies $n\leq P$ because of $b>0$. We note that
\begin{align*}
n\leq P\iff  \sum_{i=1}^n \left(\frac{x_{i}}{\rho^2}+\frac{\rho^2}{x_{i}}\right)\geq2n\iff \frac{x_{i}}{\rho^2}+\frac{\rho^2}{x_{i}} \geq2.
\end{align*}
Here, we use the inequality of arithmetic and geometric means, i.e., for any $a>0$ , $b>0$, it holds that $a+b\geq2\sqrt{ab}$. Letting $a=x^2_{i}$ and $b=\rho^4$, we have $x^2_{i}+\rho^4\ge 2\sqrt{x_i^2\rho^4}$ or equivalently, 
$(x_{i}/\rho^2)+(\rho^2/x_{i}) \ge 2$. 
Hence, it is shown that
\[d_{1}(\eta)+\frac{1}{\eta} =-n\frac{\partial}{\partial \eta}\log K_{\nu}(\eta)+\frac{a}{\eta}-P-b <0\]
for large $\eta$. From the intermediate value theorem, there exists $\eta>0$ such that $(\partial/\partial \eta)\log f(\eta)+(1/\eta)=0$.

\qed

\end{document}